\documentclass[a4paper,11pt]{article} 

\usepackage[OT1]{fontenc}
\usepackage[ansinew]{inputenc}
\usepackage{geometry}
\usepackage{latexsym}
\usepackage{amsmath}
\usepackage{amsfonts} 
\usepackage{amssymb}
\usepackage{bm}
\usepackage{textcomp}
\usepackage{graphicx}
\usepackage{adjustbox}
\usepackage{times}
\usepackage[square]{natbib}
\usepackage{deluxetable}
\usepackage[english]{babel}

\UseRawInputEncoding

\newcommand{\p}{\partial}

\begin{document}

\linespread{1}

\textbf{\large{\Large{An Exceptional \textbf{$G(2)$} Extension of the Standard Model from the Correspondence with Cayley--Dickson Algebras Automorphism Groups}}}

\noindent
\begin{center}

Nicol\`o Masi

\begin{center}
INFN \& Alma Mater Studiorum - Bologna University,

Via Irnerio 46, 40136 Bologna, Italy
\end{center}

\begin{center}
{\it masin@bo.infn.it}
\end{center}

\end{center}

\begin{abstract}

In this article I propose a new criterion to extend the Standard Model of particle physics from a straightforward algebraic conjecture: the symmetries of physical microscopic forces originate from the automorphism groups of main Cayley--Dickson algebras, from complex numbers to octonions and sedenions. This correspondence leads to a natural enlargement of the Standard Model color sector, from a $SU(3)$ gauge group to an exceptional Higgs-broken $G(2)$ group, following the octonionic automorphism relation guideline. In this picture, an additional ensemble of massive $G(2)$-gluons emerges, which is separated from the particle dynamics of the Standard Model. 
\end{abstract}

 
%


\section{Introduction} \label{Introduction} \label{Intr}

\noindent Despite its great success and prediction capability, the Standard Model (SM) of particle physics is afflicted by internal and external problems, \textit{i.e.} theoretical issues (such as hierarchy and strong CP problems) and not explained phenomena, like dark matter (DM), dark energy or matter--antimatter asymmetry \citep{ModernPP}. Above all, DM is probably the most compelling and very long-standing problem of modern physics, with no evident nor univocal solution: all the efforts made, from particle theory \citep{Bertone,ProfumoDM} to modified gravities \citep{InfraredG,ModG_largeDist,Clifton:2011jh}, have not been successful in clarifying its nature. The most convincing particle candidates, the weakly interacting massive particles (WIMPS), have not been discovered yet: direct, indirect and collider searches show no evidence of new particles approximately up to the 1 TeV scale \citep{baudis_2018,ProfumoDM,DMATLAS,DMTevish,Salvio:2019agg}. This is a strong hint that the \textit{Naturalness} criterion \citep{giudice2017dawn} for the Higgs sector and the so-called \textit{WIMP Miracle} \citep{Young:2016ala}, which postulate the existence of a thermal particle relic of the Big Bang at the electroweak scale $O(100\,$GeV$)$ which interacts via weak force, could not be a prerogative of \textit{Nature} or, at any rate, not sufficient to individuate the origin of dark matter and describe the physics beyond Standard Model. Even the possibility that the weak interaction between DM and SM particles is disfavored must be considered: new particles could hide at different energy scales and they could be not capable of interacting with the visible world, at least at the experimentally explored energies.

Therefore, to proceed in the investigation of beyond Standard Model phenomenology, one has to fill up the lack of a theoretical guideline and integrate some new simplicity criteria to select reliable candidates and explain the complex astrophysical and cosmological observations \citep{Hooper:2018kfv,Gaskins:2016cha,NicMario}.
Today physics seems to need some extra inputs to go beyond current paradigms and reach a deepest understanding 
of the dark matter conundrum: in this complex situation mathematics could provide fresh insights and conjectures to overcome physical prejudices. \\ 
Here we propose an approach based on a division algebras conjecture capable of selecting a unique extension of the SM, which introduces a branch of \textit{exceptional} matter particles from a simple and minimally high symmetry. The criterion is to identify fundamental interactions with the automorphism groups of Cayley-Dickson algebras (an automorphism is a bijective way of mapping a mathematical object to itself preserving its structure: the set of all automorphisms forms the automorphism group, \textit{i.e.} the symmetry group of the object). Then, from the automorphism of octonions (and sedenions) algebra, the promising exceptional symmetry group $G(2)$ can be pinpointed to solve the DM problem. We will demonstrate that, once broken through a Higgs-like mechanism, $G(2)$ represents the optimal gauge group to describe strong interaction and dark matter at the same time, shedding light on a primordial high energy phase transition which generated the strong sector.
This minimal extension of the SM, via Cayley--Dickson algebras automorphism correspondence, uniquely fix the content of particle physics. To the best of our knowledge, no existing work in literature is devoted to an exceptional $G(2)$ enlargement of the strong sector nor to the possibility that dark matter is formed by massive gluons from a broken-$G(2)$ gauge group, which naturally incorporates the standard $SU(3)$ color Quantum Chromodynamics (QCD): even if $G(2)$ lattice models have been largely applied to simplify standard QCD computations \citep{Wellegehausen_2011,Maas:2012ts}, the implications of such an extension of the SM have not been explored. Hence, the present dissertation is not intended as a mere review of the current status of Cayley--Dickson algebras applied to particle physics, but as a phenomenological proposal to build up an exceptional Standard Model framework and incorporate new particle physics.


\section{Fundamental forces from division algebras automorphisms}\label{sedenions}

\noindent In the last decades many attempts to connect the Standard Model of elementary 
particles with division algebras have been made, showing it is worthwhile 
establishing relations between algebraic structures and symmetry groups \citep{Book_Gursey,Book_Conway,Book_Dixon,Clifford_Lounesto,GRESNIGT2018212,Book_Maia,furey2016standard,Manogue_2010,gording2019unified}. \\
It is well-known that following the Cayley--Dickson construction process \citep{Book_Dixon,Book_Gursey}, one can build up a sequence of larger and larger algebras, adding new imaginary units. In detail, from Hurwitz and Zorn theorem \citep{Book_Conway}, one can identify the so-called division algebras $\mathbb{R}, \mathbb{C}, \mathbb{H}, \mathbb{O}$, \textit{i.e.} the only four alternative algebraic fields with no non-trivial zero divisors \citep{Baez:2001dm,Book_Dixon}, which are real numbers, complex numbers, quaternions and octonions, respectively. During the construction process, the algebras lose some peculiar properties, one at a time. For example, complex numbers are not ordered but commutative, quaternions are not commutative but associative, whereas octonions lose all the familiar commutative and associative properties, but they are still 
an alternative algebra \citep{Baez:2001dm}. The process does not terminate with octonions: applying the Cayley--Dickson construction, greater 2${}^{n}$-dimensional algebras can be constructed, for any positive integer $n$. For $n>3$, however, as anticipated, they all include non-trivial zero 
divisors, \textit{i.e.} they have problems in a general definition of norm (in abstract algebra, a non-zero element $a$ of a ring $R$ is called a zero divisor if there exists a non-zero $x$ such that $ax=0$; for general properties of zero divisors see \citep{ZeroAnnihil}). This was considered an obstacle for the use of these extended algebras, such as $n=4$ sedenions, in science. But, as shown in \citep{Gillard_2019,cawagas_2014}, sedenions should not be ruled out as playing a role in particle physics on the basis that they do not constitute a division algebra. We will return to this topic later.

The link between unitary groups and division algebras $\mathbb{A}_n$ has been diffusely studied \citep{Cacciatori:2009qu,EVANS198892,Book_Schwerdtfeger}. Unitary groups are the fundamental bricks to build the particle Standard Model, because each fundamental force can be described by a unitary or special unitary group \citep{ModernPP,PhysFSym,Super_Nath,Burgess,Schwinger}, being $G=SU(3) \times SU(2) \times U(1)$ the SM group of strong $SU(3)$, weak $SU(2)$ and hypercharge $U(1)$ interactions \citep{ModernPP}. Besides its symmetry, the SM includes three fermions families: between these three generations, particles differ by their flavour quantum number and mass, but their interactions are identical. \\ In the following, we want to briefly highlight the relations between the automorphisms of Cayley-Dickson algebras and these important physical gauge groups, including some considerations about the tripartite structure of the Standard Model. 
 
Starting from the most simple complex algebra and SM symmetry group, 
it is easy to find a direct connection between the electromagnetism (or Quantum Electrodynamics) $U(1)$ formalism and the 
complex number field $\mathbb{C}$: in fact the group $U(1)$, the smallest compact real Lie group, corresponds to the circle group $S^{1}$, consisting of all complex numbers with absolute value 1 under multiplication, which is isomorphic to the $SO(2)$ group of rotation \citep{Book_Field}. All the unitary groups contain copies of this fundamental group. 
For $n\geq1$, one can also consider for the comparison the $n$-torus $T_n$, that is defined to be $\mathbb{R}^n/\mathbb{Z}^n \cong U(n) \cong SO(2)^n \cong (S^1)^n$, where $/$ denotes the quotient group between reals and integers, which shows off the deep connection between $U(1)$ gauge symmetry and other representations strictly connected to complex numbers \citep{Book_Field,OpSym_saller}. It is also true that the $n \times n$ complex matrices which leave the scalar product $\left\langle ,\right\rangle$ invariant form the group $U(n) = Aut(\mathbb{C}^{n}, \left\langle ,\right\rangle)$, \textit{i.e.} the group of automorphisms of $\mathbb{C}^{n}$ as a Hilbert space \citep{Aguilar:1998ws}.\\
These links are not surprising because, from a mathematical point of view, the existence of infinite distinct \textit{wild} automorphisms of the complex numbers, beyond identity and complex conjugation, is well-known \citep{Clifford_Lounesto,Yale}. We find another noteworthy examination in \citep{OpSym_saller}, where the unitary group $U(1)$ is showed as defining binary complex relations $\mathbb{C} \times \mathbb{C}$, \textit{i.e.} the $U(1)$ numbers effectively operate as automorphisms of $\mathbb{C}$ via multiplication of a phase factor. As we know, the complex numbers can be expressed in polar coordinates and this implies that the general linear multiplicative group $\mathbb{C}^{*}=\mathbb{C} \backslash 0 = e^{\mathbb{C}} \cong GL(1,\mathbb{C})$ is uniquely decomposable ($e^{z}=e^{x}\cdot e^{iy}$) into the totally ordered group with real exponential $\left| \mathbb{C}^{*} \right|=e^{\mathbb{R}}$ and into the phase group with imaginary exponentials $\mathbb{C}^{*}/\left| \mathbb{C}^{*} \right|= e^{i\mathbb{R}} \cong U(1)$, which is approximately $U(1)$ (see \citep{OpSym_saller,Book_Saller} for details). This is another way to underline the intimate connection between the unitary group and the complex numbers. Furthermore, from a physical point of view, one can also think at the Riemann-Silberstein field reformulation of the electromagnetism \citep{Bialynicki_Birula_2013} in terms of a complex vector that combines the electric field $E$, as the real part, and the magnetic field $B$, as the imaginary part, in order to put in evidence this essential relation.

Even $SU(2)$ weak isospin can be clearly represented with the algebraic quaternionic basis, \textit{i.e.} \textit{Pauli matrices} \citep{Book_Maia}: $SU(2)$ naturally embeds into $\mathbb{H}$ as the group of quaternion elements of norm 1, with a perfect analogy with respect to $U(1)$ and complex numbers. More precisely, the group $SU(2)$ is isomorphic to the group of quaternions of norm 1, and it is thus diffeomorphic to the 3-sphere $S^3$ (a diffeomorphism is an isomorphism of smooth manifolds, \textit{i.e.} a map between manifolds which is differentiable and has a differentiable inverse). Indeed, since unit quaternions can be used to represent rotations in 3-dimensional space (up to a sign), there is a surjective homomorphism from $SU(2)$ to the rotation group $SO(3)$ \citep{Book_Maia} (an homomorphism is a structure-preserving map between two algebraic structures of the same type): one can show that the local $SU(2)$ spinors are exactly the same two-component spinors derived from the local quaternion matrix representation, \textit{i.e.} the three Pauli matrices along with the identity matrix $I$ (spinors are defined as vectors of a representation of the group of automorphisms of a Clifford algebra defined on space--time). In other words, the correspondence between the automorphism of quaternion algebra and the Standard Model symmetry group of weak force can be clearly shown: for quaternions $Aut(\mathbb{H})=SO(3)$, where $SO(3)$ is homomorphic to $SU(2)$ in turn, and the universal cover of $SO(3)$ is the spin group $Spin(3)$, which is isomorphic to $SU(2)$. So $SU(2)$ and $SO(3)$ algebraic structures are equivalent. An interesting demonstration of the correspondence between the two groups using $M\ddot{o}bius$ transformation is described in \citep{Moebius}. The quaternionic representation of (electro-) weak isospin has been used by many authors \citep{2012IJTP...51.3228P,Potter_2015}.\\
Hence, both in the $U(1)$ electromagnetic case and in the $SU(2)$ weak interaction, the solutions can be expressed in terms of division algebras, respectively the complex and the quaternion algebras: the division property is important to define the mathematical structure and in the determination of solutions. This could be a coincidence, but the possibility that fundamental gauge interactions can be described by the apparatus of division algebras should be explored.

It seems logical to revise the next division algebra, the octonion algebra $\mathbb{O}$ (which is not a Clifford algebra, unlike $\mathbb{R}$, $\mathbb{C}$ and $\mathbb{H}$, because non associative) \citep{Octon,Baez:2001dm} for a possible description of the $SU(3)$ gauge field \citep{Chanyal_2012,2012IJTP...51.3228P}, but the result is less clear than in quaternion case for the $SU(2)$ gauge field. The interesting fact to be considered is that the group of automorphisms of the octonion algebra, the largest of the normed division algebras, corresponds to the exceptional Lie algebra $G(2)$, the smallest among the known exceptional Lie algebras: $Aut(\mathbb{O})=G(2)$ \citep{Book_Wilson}.
So it is noteworthy to point out that the Standard Model gauge group $SU(3)$ is not isomorphic to the group of automorphisms of the octonions, which is $G(2)$. Nonetheless, it is possible to fix one of the octonion basis elements to obtain seven possible subalgebras, each of which has a subgroup of automorphisms isomorphic to $SU(3)$. For example, $SU(3)$ itself may be defined as the subgroup of $G(2)$ which leaves the octonionic unit e${}_{7}$ invariant \citep{Book_Maia}. Of course, alternative $SU(3)$ subgroups of $G(2)$ may be found, corresponding to other imaginary units.
In addition, recent works in the framework of particle physics show the possibility to rewrite \textit{Gell-Mann matrices} of $SU(3)$ strong force (the group generators) with octonions \citep{2012IJTP...51.3228P}. Also split-octonions representations have been proposed as alternative formalism for $SU(3)$ color gauge symmetry \citep{Chanyal_2012}. \\
But here a crucial difference appears: it must be noted that for $\mathbb{C}$ and $\mathbb{H}$ the direct automorphism groups contain an equal, or comparable, amount of ``mathematical information" than $U(1)$ and $SU(2)$ themselves (through the approximate algebraic correspondences, via homomorphism in $SU(2)$ case), whereas the exceptional $G(2)$ group is certainly bigger than SM $SU(3)$, as it includes $SU(3)$ and is equipped with six additional generators \citep{Holland_2003}. In other words, if we want to study the application of the octonion automorphism in physics, it is mandatory to invoke a gauge group which is not the strong color symmetry $SU(3)$.\\
Summarizing, for non real division algebras it turns out that: 
\begin{equation}\label{autom}
\mathrm{Aut}(\mathbb{C})\cong U(1),\, \mathrm{Aut}(\mathbb{H})\cong SU(2),\, \mathrm{Aut}(\mathbb{O})\equiv G(2).
\end{equation} 
These relations show an ordered correspondence between (approximate) automorphisms of algebras and gauge groups useful for Standard Model description, where $G(2)$ contains $SU(3)$ color force. We will see in the next section that, besides $SU(2)$ Pauli matrices, also $G(2)$ generators can be written in terms of $SU(3)$ Gell-Mann matrices as 14 unitary matrices. \\
Fundamental correlations between division algebras and symmetry groups, as anticipated, have been already stressed in the last decades. Using division algebras, Dixon proposed an elegant representation of particle physics in \citep{Book_Dixon}. Furey has recently suggested the appealing possibility to reformulate the SM group $G=SU(3) \times SU(2) \times U(1)$ in terms of a $\mathbb{A}=\mathbb{R}\otimes\mathbb{C}\otimes\mathbb{H}\otimes\mathbb{O}$ tensor product algebra, restarting from Dixon's work, using the concept of \textit{Ideals}, \textit{i.e.} using subspaces of proper Clifford Algebras as ``particles'' (see \citep{furey2016standard,2018IJMPA..3330005F,FUREY2015195,refId0F,PhysRevD.86.025024} for details). Also string theory and supersymmetrical theories invoked division algebra to study particle interactions \citep{Anastasiou_2014,QFT4Math_Deligne,Book_Superstring,EVANS198892,Kugo:1982bn,Preitschopf:1995qf}. Moreover, $G(2)$ as automorphism of $\mathbb{O}$  has important applications in terms of the so-called $G_2$ structure or $G_2$ manifolds \citep{g2manifold}, in the context of M-theory \citep{Becker_2015}.
Indeed, one solid reason for studying division algebras in relation to particle symmetries is that, unlike Lie algebras and Clifford algebras \citep{Clifford_Lounesto}, there is a finite number of division algebras and corresponding automorphisms (see again the extensive works of Dixon \citep{Book_Dixon}). If we start with a division algebra, the physical symmetries are dictated by the mathematical structure and the choice of a proper symmetry group is constrained.\\
To proceed with the reasoning, we are going to see why 16-dimensional sedenions can be easily added to this picture and how dark matter description can benefits from this algebraic facts, summarizing the main features of sedenions algebra.

The sedenion algebra is the fifth Cayley-Dickson algebra $\mathbb{A}_4=\mathbb{S}$, where $\mathbb{A}_{0,1,2,3}$ correspond to reals, complex numbers, quaternions and octonions. This is not a division algebra, it is non-commutative, non-associative, and non-alternative (an algebra $A$ is alternative if the subalgebra generated by any two elements is associative, \textit{i.e.} iff for all $a,b \in A$ we have $(aa)b = a(ab), (ba)a = b(aa)$ \citep{culbert_2007}), hence it cannot be a composition algebra \citep{SMITH1995128,Gillard_2019} (where a composition algebra is an algebra $A$ over a field $K$ with a non-degenerate quadratic form $N$, called \textit{norm}, that satifies $N(ab) = N(a)N(b)$ for all $a,b$ in $A$ \citep{Book_Conway}). However sedenion algebra is power-associative and flexible (an algebra is power-associative if the subalgebra generated by any one element is associative: it is a sort of lowest level of associativity \citep{Baez:2001dm}; the flexible property, for any $a,b \in A$, can be defined as $a(ba) = (ab)a$), and satisfies the weak inversive properties for non-zero elements. Each Cayley-Dickson algebra satisfies the weak inversive property:  $a^{-1}(ab)=a(a^{-1}b), (ba^{-1})a=(ba)a^{-1}, a^{-1}(ab)=(ba)a^{-1}$ -- see \citep{culbert_2007,cawagas_2014} for details. In principle, the Cayley-Dickson construction can be indefinitely carried on and, at each step, a new power-associative and flexible algebra is produced, doubling in size. So, in first approximation, no new fundamental properties and information are added nor lost enlarging the algebra beyond sedenions.
One can choose a canonical basis for $\mathbb{S}$ to be $E_{16}=\lbrace e_i\in\mathbb{S}\vert i=0,1,...,15\rbrace$ where $e_0$ is the real unit and $e_1,...,e_{15}$ are anticommuting imaginary units. In this basis, a general element $A\in\mathbb{S}$ is written as
\begin{eqnarray}
A= \sum_{i=0}^{15} a_i  e_i= a_0+\sum_{i=1}^{15} a_i  e_i,\qquad a_i\in\mathbb{R}.
\end{eqnarray}
The basis elements satisfy the multiplication rules
\begin{eqnarray}
\nonumber e_0&=&1,\qquad e_0e_i=e_ie_0=e_i,\\
e_i^2&=&-e_0, \qquad i\neq0,\\
\nonumber e_ie_j&=&\gamma^k_{ij}e_k \qquad i\neq0, \qquad i\neq j,
\end{eqnarray}
with $\gamma^k_{ij}$ the real structure constants, which are completely antisymmetric. For two sedenions $A,B$, one has
\begin{eqnarray}
AB=\left( \sum_{i=0}^{15} a_i  e_i \right) \left( \sum_{i=0}^{15} b_j  e_i\right) =\sum_{i,j=0}^{15} a_ib_j  (e_ie_j)=\sum_{i,j,k=0}^{15} f_{ij}\gamma^k_{ij}  e_k,
\end{eqnarray}
where $f_{ij}\equiv a_ib_j$. 

Because the sedenion algebra is not a division algebra, it contains zero divisors: for $\mathbb{S}$ these are elements of the form
\begin{eqnarray}
(e_a+e_b)\circ(e_c+e_b)=0,\qquad e_a,e_b,e_c,e_d\in\mathbb{S}.
\end{eqnarray}
There are 84 such zero divisors in sedenion space and the subspace of zero divisors with unit norm is homeomorphic to $G(2)$ \citep{moreno1997zero,ZeroAnnihil}.
To understand the role and emergency of zero divisors, one has to consider not only single algebras but also compositions of them.
For example, whereas $\mathbb{R},\mathbb{C},\mathbb{H}$ and $\mathbb{O}$ are by themselves division algebras, their tensor products, such as $\mathbb{C}\otimes\mathbb{H}$, $\mathbb{C}\otimes\mathbb{O}$ and $\mathbb{R}\otimes\mathbb{C}\otimes\mathbb{H}\otimes\mathbb{O}$, largely applied in SM algebraic extensions, are not, and in fact the zero divisors of these algebras play a crucial role in the construction of Furey's Ideals \citep{FUREY2015195,furey2016standard,2018IJMPA..3330005F,refId0F}. Moreover, the two by two compositions of division algebras, which are not division algebras and contain zero divisors, are the subjects of the well-known \textit{Freudenthal--Tits magic square} \citep{barton2000magic,BARTON2003596,Cacciatori:2012cb}:
\begin{center}
\begin{tabular}{|l|c|c|c|c|}
\hline
\multicolumn{1}{|c|}{$\otimes$} & $\mathbb{R}$ & $\mathbb{C}$ & $\mathbb{H}$ & $\mathbb{O}$ \\
\hline
$\mathbb{R}$  &    $SO(3)$   &   $SU(3)$       & $Sp(3)$    & $F_4$ \\ 
$\mathbb{C}$  &    $SU(3)$   &  $SU(3)^2$      & $SU(6)$    & $E_6$ \\ 
$\mathbb{H}$  &    $Sp(3)$   &   $SU(6)$       & $SO(12)$   & $E_7$ \\ 
$\mathbb{O}$  &    $F_{4}$   &   $E_6$         & $E_7$      & $E_8$ \\
\hline
\end{tabular}
\end{center}

a symmetric square ($SO(N)$ and $SU(N)$ are the usual special orthogonal and unitary groups of order $N$, $Sp(3)$ is the symplectic group of order three), which exhibits the ``unexpected" relation between octonions products and exceptional groups ($F_4$, $E_6$, $E_7$, $E_8$) \citep{Book_Wilson}, except for the exceptional $G(2)$ which represents octonions automorphism itself. The exceptional groups on the last line/row are not exactly automorphisms of the octonions products, because of mathematical problems in the definition of projective planes, due to the appearance of zero divisors: they are called bioctonions ($\mathbb{C}\otimes\mathbb{O}$),  quateroctonions ($\mathbb{H}\otimes\mathbb{O}$) and octooctonions ($\mathbb{O}\otimes\mathbb{O}$) and find correspondence into Jordan's algebras \cite{Baez:2001dm}. Exceptional $E_i$ are also largely used in supergravity and string theory \citep{Book_Superstring,Super_Nath}.
Therefore it seems reasonable to continue the Cayley-Dickson algebraic construction into the non-division algebras, such as $\mathbb{S}$.\\
Interestingly, in \citep{gillard2019cell8,Gillard_2019} the authors put in evidence an important relation between sedenions and the exceptional group $G(2)$, demonstrated by Brown in \citep{Brown67}:
\begin{eqnarray}\label{sed}
\mathrm{Aut}(\mathbb{S})=\mathrm{Aut}(\mathbb{O})\times S_3.
\end{eqnarray}
where we know that $\mathrm{Aut}(\mathbb{O})=G(2)$ and $S_3$ is the permutation group of degree three. So the inner symmetries of this non-division algebra can be again extracted from the automorphism group of octonions and, in particular, from a proper product of the exceptional $G(2)$ group with a symmetric group. The only difference between octonions and sedenions automorphism groups is a factor of the permutation group $S_3$: this permutation group can be constructed from the \textit{triality} automorphism of the spin group $Spin(8)$ (triality is a trilinear map among three vector spaces, most commonly described as a special symmetry between vectors and spinors in 8-dimensional euclidean space -- see \citep{Clifford_Lounesto,Baez:2001dm} for details). Eq.(\ref{sed}) suggests that the fundamental symmetries of $\mathbb{S}$ are the same as those of $\mathbb{O}$, even if the factor $S_3$ introduces a three copies scenario, that is exactly what we need in order to describe the observed three generations of fermions in the Standard Model of particles.\\
The previous formula can be generalized, for an arbitrary algebra constructed via Cayley-Dickson process (for $n>3$), into \citep{Gillard_2019,EakinAutomorph}
\begin{eqnarray}
\mathrm{Aut}(\mathbb{A}_n)\cong\mathrm{Aut}(\mathbb{O})\times (n-3)S_3.
\end{eqnarray}
This tells us that the underlying symmetry is always $G(2)$, the automorphism group of the octonions. The higher Cayley-Dickson algebras only add additional \textit{trialities}, \textit{i.e.} copies of $G(2)$, and reasonably no new physics beyond sedenions. Futhermore, sedenion algebra might represent the archetype of all non-associative and non-division flexible algebras, if $n>3$ Cayley-Dickson algebras do not differ from sedenions for what concerns the multilinear identities (or algebraic properties) content, as suggested in \citep{triginta}.\\
In this picture, sedenion algebra could constitute the searched simplicity criterion to select the full symmetry of a three generations Standard Model strong force and include a new particle physics content, which might represent the unknown dark matter sector. 
This could be also read as a sort of a \textit{naive} indirect proof that fundamental forces should be a small number (only three), because all algebras beyond octonions point towards the very same exceptional group, adding only copies (particle generations).
Finally, as it will be discussed in the next section, to recover the usual $SU(3)$ strong force the octonions-sedenions automorphism group must be broken at our energy scales and new physics extracted: this enlarged algebraic content is going to be associated to dark matter.\\  
\noindent So, without the presumption of a rigorous and definitive mathematical definition of the problem, we can reformulate and summarize the algebraic phenomenological conjecture in a general way as follows.

\noindent The fundamental symmetry of the Standard Model of particle physics with three fermion families might be 
the realization of some tensor products between the associative division algebras and the most comprehensive non-division algebra 
obtained through the Cayley--Dickson construction, \textit{i.e.} the sedenion algebra. The sedenionic description, like the octonionic one, corresponds, via automorphism, to the simplest exceptional group $G(2)$, but tripled. It could provide an explanation to the $N=3$ fermion families of the Standard Model, which lie in the sedenions $S_3$ automorphism factor, as suggested by \citep{Gillard_2019}. This is consistent with the proposal of a $S_3$--invariant extension of the Standard Model, as discussed in \citep{Kubo_2004,Kubo_2005,Mondrag_n_2007,Gonz_lez_Canales_2012}.\\
\noindent The gauge groups $U(1), SU(2), SU(3)$, describing the three fundamental forces, find mathematical correspondence into the division algebras $\mathbb{C},\mathbb{H},\mathbb{O}$ respectively: Table \ref{Tab:All} summarizes this correspondence.
However, whereas $U(1)$ and $SU(2)$ are approximate isomorphisms of complex and quaternion algebras automorphism groups (see Eq.(\ref{autom})), the octonion and sedenion automorphism relations point towards a different group, which is manifestly larger than the usual 8-dimensional $SU(3)$ color group of the Standard Model, \textit{i.e.} the 14-dimensional $G(2)$ group; $SU(3)$ and $G(2)$ differ for 6 dimensions/generators. Therefore 
\begin{equation}
 \begin{aligned}
\mathrm{Aut}(\mathbb{C})\times \mathrm{Aut}(\mathbb{H})\times \mathrm{Aut}(\mathbb{S}) = \mathrm{Aut}(\mathbb{C})\times \mathrm{Aut}(\mathbb{H})\times \mathrm{Aut}(\mathbb{O})\times S_3 = \\ 
U(1) \times SU(2) \times G(2) \times S_3
 \end{aligned}
\end{equation} 
could give the overall unbroken Standard Model symmetry. This is the first main statement of the present dissertation.
Here the automorphism selection is invoked to predict something beyond current SM, and $SU(3)$ in particular, and it works as a guideline to replace $SU(3)$ color itself with the smallest exceptional group: \textit{fundamental forces must be isomorphic to the automorphisms groups of the division algebras built up through the Cayley--Dickson construction.} Tensor products between the corresponding algebras (see Freudenthal--Tits magic square) could be effective symmetries but not fundamental forces.\\
A new particle content come from the aforementioned difference between $G(2)$ and $SU(3)$ groups and lie in the spectrum gap between them. Following the Cayley--Dickson algebraic automorphism criterion, no more physics is needed nor predicted, except for the six additional degrees of freedom, \textit{i.e.} boson fields, which represent the discrepancy between $G(2)$ and $SU(3)$ generators. Hence, the automorphism selection rule extends the strong color sector and provides a rich exceptional phenomenology.

\begin{table}
\begin{center}
\small\addtolength{\tabcolsep}{-5pt}
\scalebox{0.82}{
\begin{tabular}[hb!]{|c|c|c|c|c|c|c|c|c|c|c|}
\hline
\multicolumn{1}{|c|}{Charge ($n_g$)} & Group & Force & Algebra & Dim & Commutative & Associative & Alternative & Normed & Flexible \\
\hline
$Q (1)$          &    $U(1)$       &   EM         	& $\mathbb{C}$               & 2      &  Yes&Yes &Yes &Yes &Yes \\ 
$T (3)$         &    $SU(2)$       &   Weak      		& $\mathbb{H}$               & 4      &  No &Yes &Yes &Yes &Yes \\ 
$C (8)$         &    $SU(3)$       &   Strong   		& $\mathbb{O}$ or $\mathbb{S}$   & 8/16   &  No &No  &Yes &Yes &Yes \\ 
$EC(6)$         &  broken-$G(2)$   &   Exceptional Strong  & $\mathbb{O}$ or $\mathbb{S}$   & 8/16   & No  &No  &No  &No  &Yes  \\ 
\hline
\end{tabular}}
\end{center}
\caption{\footnotesize{Schematic correspondence between forces, groups and algebras. In the first column the charge of the physical interaction is displayed along with the number $n_g$ of associated generators (bosons). $Q,T,C$ are usual SM electric charge, weak isospin and color charge, respectively; here \textit{EC} stands for ``exceptional--colored", to indicate the six broken generators which originate the massive exceptional $G(2)$ bosons which have quark and anti-quark color quantum numbers (see next section). The second and third columns associate gauge groups and forces, highlighting the link between $G(2)$ and the 6 new exceptional--colored particles, separated from visible strong phenomena. $G(2)$ algebraic automorphism representation is valid for both octonions and sedenions (the only difference is the $S_3$ factor). In principle, strong force and exceptional sector represent the same interaction but they are disconnected, coming from the broken exceptional symmetry. For this reason their algebras are both displayed as $\mathbb{O}$ or $\mathbb{S}$. Algebraic dimensions are showed in the fifth column. As shown in the subsequent columns, each division algebra loses inner properties hierarchically, from commutativity to alternativity, as the dimensions increase. All algebras are flexible (and power-associative). See \citep{SMITH1995128,Baez:2001dm} for proper descriptions of the algebraic properties and insights.}}
\label{Tab:All}
\end{table}

A further novelty is the definition of an original algebraic criterion to predict physics beyond the Standard Model, which substitutes Higgs Naturalness and the Wimp Miracle. In this scenario, the strong force acquires a more complex structure, which includes the usual color sector and an enlarged strong exceptional dynamics, due to six residual generators of exceptional $G(2)$, which might gain mass via a symmetry breaking: to recover standard $SU(3)$ color strong force description, the new $G(2)$ color sector should be broken by a Higgs-like mechanism and separated into two parts, one visible and the other excluded from the dynamics due to its peculiar properties.\\
The next section is devoted to a deep analysis of the exceptional $G(2)$ group and to the emergency of these massive exceptional bosons.

\section{A $\textbf{G(2)}$ gauge theory for the strong sector}\label{G2 QCD}

\noindent $G(2)$ can be described as the automorphism group of the octonion algebra or, equivalently, as the subgroup of the special orthogonal group $SO(7)$ that preserves any chosen particular vector in its 8-dimensional real spinor representation \citep{Octon,Book_Aschbacher,Clifford_Lounesto}. The group $G(2)$ is the simplest among the exceptional Lie groups \citep{Cacciatori:2009qu}; it is well known that the compact simple Lie groups are completely described by the following classes: $A_N(=SU(N+1)),  B_N(=SO(N+1)),  C_N(=Sp(N)),  D_N(=SO(2N))$ and exceptional groups $G_2,  F_4,  E_6,  E_7,  E_8$, with $N=1,2,3,...$(for $D_N$, $N>2$) \citep{AdvModAlg}. Among them, only $SU(2), SU(3), SO(4)$ and symplectic $Sp(1)$ have 3-dimensional irreducible representations and only one, $SU(3)$, has a complex triplet representation (this was one of the historical criteria to associate $SU(3)$ to the three color strong force, with quark states different from antiquarks states \citep{FondQCD}). There is only one non-Abelian simple compact Lie algebra of rank 1, \textit{i.e.} the one of $SO(3) \simeq SU(2) = Sp(1)$, which describes the weak force, whereas there are four of rank 2, which generate the groups $G(2)$, $SO(5) \simeq Sp(2)$, $SU(3)$ and $SO(4) \simeq SU(2) \otimes SU(2)$, with 14, 10, 8 and 6 generators, respectively \citep{Holland_2003}. \\
If we want to enlarge the QCD sector to include dark matter, it is straightforward we have to choose $G(2)$ or $SO(5)$. The group $G(2)$, beside its clear relation with division algebras described in the previous section, is of particular interest because it has a trivial center, the identity, and it is its own universal covering group, meanwhile $SO(5)$ has $\mathbb{Z}_2$ as a center (and $SU(3)$ has $\mathbb{Z}_3$); $SO(N)$ in general are not simply connected and their universal covering groups for $n>2$ are spin $Spin(N)$ \citep{Pepe_2006}. It is also well-know in literature that $G(2)$, thanks to its aforementioned peculiarities, can be used to mimic QCD in lattice simulations, avoiding the so-called sign problem \citep{Wipf:2013vp} which afflicts $SU(3)$. Proposing to enlarge QCD above the TeV scale and have the SM as a low energy theory is surely not an unprecedented nor odd idea: for example, modern composite Higgs theories \citep{Marzocca_2014,Da_Rold_2019,Cacciapaglia_2019} try to introduce (cosets) gauge groups beyond $SU(3)$, such as $SU(6)/SO(6)$, $SO(7)/SO(6)$ or $SO(5)/SO(4)$, dealing with multiple Higgs, strong composite states and dark matter candidates.

Focusing on the present proposal, $G(2)$ can be constructed as a subgroup of $SO(7)$, which has rank 3 and 21 generators \citep{Holland_2003,Pepe_2006}. The $7 \times 7$ real matrices $U$ of the group $SO(7)$ have determinant 1, orthogonal relation $UU^\dag=1$ and fulfill the constraint $U_{ab} U_{ac} = \delta_{bc}$.
The $G(2)$ subgroup is described by the matrices that also satisfy the cubic constraint
\begin{equation}
\label{cubic}
T_{abc} = T_{def} U_{da} U_{eb} U_{fc}
\end{equation}
where $T$ is an anti-symmetric tensor defining the octonions multiplication rules, whose non-zero elements are
\begin{equation}
\label{structure}
T_{127} = T_{154} = T_{163} = T_{235} = T_{264} = T_{374} = T_{576} = 1.
\end{equation}
To explicitly construct the matrices in the fundamental representation, one can choose the first eight generators of $G(2)$ as \citep{Holland_2003,Pepe_2006}:
\begin{equation}
\label{su3gen}
\Lambda_a = \frac{1}{\sqrt{2}} \left( \begin{array}{ccc} \lambda_a & 0 & 0 \\ 
																													0 & \; -\lambda_a^* & 0\\ 0 & 0 &  0\
\end{array} \right).
\end{equation}
where $\lambda_a$ (with $a \in \{1,2,...,8\}$) are the Gell-Mann generators of $SU(3)$, which indeed is a subgroup of $G(2)$, with standard normalization $\mbox{Tr} \lambda_a \lambda_b = \mbox{Tr} \Lambda_a \Lambda_b = 2 \delta_{ab}$. $\Lambda_3$ and $\Lambda_8$ are diagonal and represent the Cartan generators w.r.t. $SU(3)$. The $G(2)$ coset space by its subgroup $SU(3)$ is a 6-sphere $G(2)/SU(3)\cong S^6\cong SO(7)/SO(6)$ \citep{Coset}, in analogy with the composite Higgs proposal \citep{Da_Rold_2019}.\\ 
The remaining six generators can be found studying the root and weight diagrams of the group \citep{RevModPhys.34.1,Carone_2008,PhysRevD.99.116024}, and can be written as:
\begin{equation}
\Lambda_{9}=\frac{1}{\sqrt{6}} 
\left(\begin{array}{ccc}0&-i\lambda_2&\sqrt{2}e_3 \\i\lambda_2&0&\sqrt{2}e_3 
\\ \sqrt{2}e_3^T & \sqrt{2}e_3^T&0\end{array}\right),
\Lambda_{10}=\frac{1}{\sqrt{6}} \left(\begin{array}{ccc}0&-\lambda_2&i\sqrt{2}e_3 
\\-\lambda_2&0&-i\sqrt{2}e_3 \\ -i\sqrt{2}e_3^T & i\sqrt{2}e_3^T&0\end{array}\right),
\end{equation}
\begin{equation}
\Lambda_{11}=\frac{1}{\sqrt{6}} 
\left(\begin{array}{ccc}0&i\lambda_5&\sqrt{2}e_2 \\-i\lambda_5&0&\sqrt{2}e_2 
\\ \sqrt{2}e_2^T & \sqrt{2}e_2^T&0\end{array}\right),
\Lambda_{12}=\frac{1}{\sqrt{6}} 
\left(\begin{array}{ccc}0&\lambda_5&i\sqrt{2}e_2 \\ \lambda_5&0&-i\sqrt{2}e_2 
\\ -i\sqrt{2}e_2^T & i\sqrt{2}e_2^T&0\end{array}\right),
\end{equation}
\begin{equation}
\Lambda_{13}=\frac{1}{\sqrt{6}} 
\left(\begin{array}{ccc}0&-i\lambda_7&\sqrt{2}e_1 \\i\lambda_7&0&\sqrt{2}e_1 
\\ \sqrt{2}e_1^T & \sqrt{2}e_1^T&0\end{array}\right),
\Lambda_{14}=\frac{1}{\sqrt{6}} \left(\begin{array}{ccc}0&-\lambda_7&i\sqrt{2}e_1 
\\-\lambda_7&0&-i\sqrt{2}e_1 
\\ -i\sqrt{2}e_1^T & i\sqrt{2}e_1^T&0\end{array}\right),
\end{equation}
where $e_i$ are the unit vectors
\begin{equation}
e_1= \left(\begin{array}{c} 1 \\ 0 \\ 0\end{array}\right),
\,\,\,\,\,
e_2= \left(\begin{array}{c} 0 \\ 1 \\ 0\end{array}\right),
\,\,\,\,\,
e_3= \left(\begin{array}{c} 0 \\ 0 \\ 1\end{array}\right)\,\,\,.
\end{equation}
In the chosen basis of the generators it is manifest that, under $SU(3)$ subgroup transformations, the 7-dimensional representation decomposes into \citep{Holland_2003,Pepe_2006}
\begin{equation}
\label{dec7}
\{7\} = \{3\} \oplus \{\overline 3\} \oplus \{1\} .
\end{equation}
Since all $G(2)$ representations are real, the $\{7\}$ representation is identical to its complex conjugate, so that $G(2)$ ``quarks'' and ``anti-quarks'' are conceptually indistinguishable. This representation describes a $SU(3)$ quark $\{3\}$, a $SU(3)$ anti-quark $\{\overline 3\}$ and a $SU(3)$ singlet $\{1\}$. 
The generators transform under the 14-dimensional adjoint representation of $G(2)$ \citep{Holland_2003,Pepe_2006}, which decomposes into \citep{Holland_2003,Pepe_2006,Pepe_2007}
\begin{equation}
\label{dec14}
\{14\} = \{8\} \oplus \{3\} \oplus \{\overline 3\}.
\end{equation}
So the $G(2)$ ``gluons'' ensemble is made of $SU(3)$ gluons $\{8\}$ plus six additional ``gluons'' which have $SU(3)$ quark and 
anti-quark color quantum numbers. As mentioned before, the center of $G(2)$ is trivial, containing only the identity, and the universal covering group of $G(2)$ is $G(2)$ itself. This has important consequences for confinement \citep{Pepe_2006,Pepe_2007,Greensite_2007,Nejad_2014}: we will see that the color string between $G(2)$ ``quarks'' is capable of breaking via the creation of dynamical gluons. As discussed in \citep{Holland_2003}, the product of two fundamental representations
\begin{equation}
\label{diq}
\{7\} \otimes \{7\} = \{1\} \oplus \{7\} \oplus \{14\} \oplus \{27\}, 
\end{equation}
shows a singlet $\{1\}$: as a noteworthy implication, two $G(2)$ ``quarks'' can form a color-singlet, or a ``diquark''. Moreover, just as for $SU(3)$ color, three $G(2)$ ``quarks'' can form a color-singlet ``baryon'':
\begin{equation}
\label{prod}
\{7\} \otimes \{7\} \otimes \{7\} = \{1\} \oplus 4 \; \{7\} \oplus 
2 \; \{14\} \oplus 3 \; \{27\} \oplus 2 \; \{64\} \oplus \{77\}.
\end{equation}
Due to the fact that ``quarks'' and ``antiquarks'' are indistinguishable, it is straightforward to show for the one flavor $N_f=1$ case that the $U(1)_{L=R} = U(1)_B$ baryon number symmetry of $SU(3)$ QCD is reduced to a ${Z_{2}}_{B}$ symmetry \citep{Holland_2003,Wellegehausen:2011jz}: one can only distinguish between states with an even and odd number of ``quark'' constituents. \\
Another useful example is
\begin{eqnarray}\label{hybrid}
&&\{7\} \otimes \{14\}\otimes \{14\}\otimes \{14\} = \{1\} \oplus ...
\end{eqnarray}
From this composition it is clear that three $G(2)$ ``gluons" are sufficient to screen a $G(2)$ ``quark", producing a color-singlet hybrid $qGGG$.
It is also true that:
\begin{equation}
\label{quark4}
\{7\} \otimes \{7\} \otimes \{7\} \otimes \{7\}= 4\{1\} \oplus ...
\end{equation}
so that the product contains four singlets.\\
Summarizing: a $G(2)$ gauge theory has colors, anticolors and color-singlet, and 14 generators. So it is characterized by 14 gluons, 8 of them transforming as ordinary gluons (as an octuplet of $SU(3)$), while the other 6 $G(2)$ gauge bosons separates into $\{3\}$ and $\{\overline 3 \}$, keeping the color quarks/antiquarks quantum numbers, but they are still vector bosons. A general Lagrangian for $G(2)$ Yang-Mills theory can be written as \citep{ModernPP,Holland_2003,Pepe_2006}:
\begin{equation}
{\cal L}_{YM}[A] = -\frac{1}{2} \mbox{Tr} [F_{\mu\nu}^2],
\end{equation}
with the field strength
\begin{equation}
F_{\mu\nu} = \p_\mu A_\nu - \p_\nu A_\mu - i g_G [A_\mu,A_\nu],
\end{equation}
obtained from the vector potential
\begin{equation}
A_\mu(x) = A_\mu^a(x) \frac{\Lambda_a}{2}.
\end{equation}
with $g_G$ a proper coupling constant for all the gauge bosons and $\Lambda_a$ the $G(2)$ generators. 
The Lagrangian is invariant under non-Abelian gauge transformations $A_\mu' = U (A_\mu + \p_\mu) U^\dagger$, with $U(x) \in G(2)$. $G(2)$ Yang-Mills theory is asymptotically free like all non-Abelian $SU(N)$ gauge theories and, on the other hand, we expect confinement at low energies \citep{Pepe_2006}.
The $G(2)$ confinement is surely peculiar with a different realization with respect to $SU(3)$, where gluons cannot screen quarks (and screening arises due to dynamical quark-antiquark pair creation). In particular, as we have already seen in Eq.(\ref{hybrid}), $G(2)$ admits a new form of exceptional confinement. It has been showed that $G(2)$ lattice Yang-Mills theory is indeed in the confined phase in the strong coupling limit \citep{Holland_2003}.\\
But we know that $G(2)$ is not a proper gauge theory for a real Quantum Chromodynamics theory. Therefore we must add a Higgs-like field in the fundamental $\{7\}$ representation in order to break $G(2)$ down to $SU(3)$. The consequence is simple and fundamental: 6 of the 14 $G(2)$ ``gluons'' gain a mass proportional to the vacuum expectation value (vev) $w$ of the Higgs-like field, the other 8 $SU(3)$ gluons remaining untouched and massless. 
The Lagrangian of such a $G(2)$-Higgs model can be written as \citep{Maas:2012ts,Holland_2003,Pepe_2006,Pepe_2007}:
\begin{equation}
{\cal L}_{G_{2}H}[A,\Phi] = {\cal L}_{YM}[A] + (D_\mu \Phi)^2 - V(\Phi)
\end{equation}
where $\Phi(x) = (\Phi^1(x),\Phi^2(x),...,\Phi^7(x))$ is the real-valued Higgs-like field, $D_\mu \Phi = (\p_\mu - i g_G A_\mu) \Phi$ is the covariant derivative and 
\begin{equation}\label{Hpot}
V(\Phi) = \lambda (\Phi^2 - w^2)^2 
\end{equation}
the quadratic scalar potential, with $\lambda > 0$. Because of the  $\{7\}\otimes \{7\}\otimes \{7\} = \{1\} \oplus ...$ singlet state seen before, in the fundamental representation a Higgs cubic term should be considered but, according to the antisymmetric property of $T_{abc}$, such a term disappears. Following the product in Eq.(\ref{quark4}), the four singlets corresponds to $w^2\Phi^2$, $\Phi^4$ and two vanishing due to antisymmetry, making the aforementioned potential general and consistent with $G(2)$ symmetry breaking and renormalizability. We can choose a simple vev like $\Phi_0=\frac{1}{\sqrt{2}}(0,0,0,0,0,0,w)$ to break $G(2)$ and re-obtain the familiar unbroken $SU(3)$ symmetry: it is easy to notice from the diagonal and non-diagonal structure of Eq. (11-14) that
\begin{eqnarray}
\Lambda_{1-8} \Phi_0 = 0  & \textrm{(unbroken generators)}\\
\Lambda_{9-14} \Phi_0 \neq 0 & \textrm{(broken generators)}
\end{eqnarray}
Plugging this scalar field vev into the square of the Higgs covariant derivative, we get the usual quadratic term in the gauge fields 
\begin{equation}\label{mmass}
g_G^2 \Phi_0^\dag \frac{\Lambda_a}{2} \frac{\Lambda_b}{2} \Phi_0 A_\mu^a(x) A^{\mu,b}(x) = \frac{1}{2} M_{ab} A_\mu^a(x) A^{\mu,b}(x)
\end{equation}
that gives the diagonal mass matrix $M_{ab}$ for the gauge bosons, of which we can use the aforementioned trace normalization relation, which Gell-Mann matrices and $G(2)$ generators share, to put the squared masses terms $g_G^2 w^2$ in evidence. 
This new scalar $\Phi$, which acquires a typical mass of 
\begin{equation}\label{mH}
M_H = \sqrt{2\lambda} w
\end{equation}
from the expansion of the potential about its minimum \citep{ModernPP}, should be a different Higgs field w.r.t. the SM one, with a much higher vev, in order to disjoin massive gluons dynamics from SM one, and a strong phenomenology. Such a strongly coupled massive field could be ruled out by future LHC and Future Circular Collider searches \citep{Adhikary_2019} (it is enough to think of heavy scalars models searches, such as the two-Higgs doublet model \citep{Arhrib_2014} or the composite Higgs models \citep{Banerjee_2018}).
In this picture, as anticipated, following the standard Higgs mechanism to build up the dark candidates, 6 massless Goldstone bosons are eaten and become the longitudinal components of $G(2)$ vector gluons corresponding to the broken generators, which acquire the eigenvalue mass 
\begin{equation}\label{mGG}
M_G = g_G w
\end{equation}
through the Higgs mechanism \citep{ModernPP}, according to Eg. (\ref{mmass}), and exhibit the color quarks/antiquarks quantum numbers. No additional Yukawa-like terms are needed for the purpose of the present proposal, so that quarks remain massless at the scale of $G(2)$ symmetry breaking, since the SM Higgs has not yet acquired its vev. 
Then, if the sedenions description via automorphisms group is invoked, the symmetry breaking process could in principle act on three different copies of $G(2)$, expressed by the permutation factor $S_3$ which keeps track of the three fermion families. In other words, a Higgs sector (the SM one or an additional strong--coupled one for $G(2)$) of a $S_3$--invariant extension of the SM could also break the flavour symmetry in order to produce the correct patterns of different masses and mixing angles for fermions families (see \citep{Kubo_2004,Gonz_lez_Canales_2013} for insights). Additionally, it has been shown that, in a phenomenologically viable electroweak $S_3$ extension of the SM, $S_3$ symmetry should be broken to prevent flavor changing neutral currents \citep{Kubo_2004} and the Higgs potential becomes more complicated due to the presence of three Higgs fields \citep{Mondrag_n_2007}. For simplicity, we could assume that this hypothetical process, involving $S_3$ breaking and Yukawa fermion masses generation, triggers at the electroweak scale, without interfering with the $G(2)$ Higgs potential.\\ 
Using the Higgs mechanism to smoothly interpolate between $SU(3)$ and $G(2)$ Yang-Mills theory, we can study the deconfinement phase transition. 
In the $SU(3)$ case this transition is weakly first order. In fact, in (3 + 1) dimensions only $SU(2)$ Yang-Mills theory manifest a second order phase transition, whereas, in general, $SU(N)$ Yang-Mills theories with any higher $N$ seem to have first order deconfinement phase transitions \citep{Lucini_2004,Nada:2015aia,Lucini_2014,Teper:2009uf,Panero_2009}, which are more markedly first order for increasing $N$. The peculiarities of the phase transition from lattice $G(2)$-Higgs to $SU(3)$ have been extensively studied in \citep{Wellegehausen_2011,Nejad_2014,Cossu_2007,vonSmekal:2013qqa,thermoG22014}, confirming that $G(2)$ gauge theory has a finite-temperature deconfining phase transition mainly of first order and a similar but discernable behavior with respect to $SU(N)$ \citep{thermoG22014}. It is interesting to mention that it has been shown \citep{Cutting_2018,PhysRevLett.115.181101,Zhou_2020,2021JHEP...05..160Z} that first order phase transitions in the early Universe could produce gravitational waves detectable by future space--based gravitational observatories such as LISA.

Moving back to the $G(2)$ color string, the breaking of this string between two static $G(2)$ ``quarks" happens due to the production of two triplets of $G(2)$ ``gluons" which screen the quarks. Hence, the string breaking scale is related to the mass of the six $G(2)$ ``gluons" popping out of the vacuum. The resulting quark-gluons bound states (colorless $qGGG$ states) coming from the string breaking, must be both $G(2)$-singlets and $SU(3)$-singlets. When we switch on the interaction with the Higgs field, six $G(2)$ gluons acquire a mass thanks to the Higgs mechanism. The larger is $M_G$, the greater is the distance where string breaking occurs. When the expectation value of the Higgs-like field is sent to infinity, so that the 6 massive $G(2)$ ``gluons" are completely removed from the dynamics, also the string breaking scale is infinite. Thus the scenario of the usual $SU(3)$ string potential reappears. For small $w$ (on the order of $\Lambda_{QCD}$), on the other hand, the additional $G(2)$ ``gluons'' could be light and participate in the dynamics. As long as $w$ remains finite, as we know it should be in the SM and in its extensions, the massive $G(2)$ ``gluons'' can mediate weak baryon number violating processes \citep{Holland_2003} (only in the $w \rightarrow \infty$ limit baryon number is an exact discrete symmetry of the Lagrangian). Finally, for $w = 0$ the Higgs mechanism disappears and we come back to $G(2)$. As stressed before, hereafter only high $w$ values (with $w$ much greater that the SM Higgs vev) are considered in order to realize a consistent dark matter scenario.\\

For what concerns the hadronic spectrum of a hypothetical $G(2)$-QCD, the physics appears to be qualitatively similar to $SU(3)$ QCD \citep{masses2014}, but richer. This can be easily demonstrated from the decomposition of representations products, like Eq.(\ref{diq}), (\ref{prod}), (\ref{hybrid}), (\ref{quark4}). In the (massless) spectrum of the unbroken $G(2)$ phase there are many more states beyond standard mesons and baryons: one-quark-three-$G(2)$ gluons hybrid states (and, in general, the quark confinement for one-quark-$N$-$G(2)$ gluons, with $N\geq 3$), diquarks, $(qqqq)$ tetraquarks and $(qqqqq)$ pentaquarks. States with baryon number 0 and 3 are in common with QCD whereas $n_B=1,2$, of $J=1/2$ hybrids and $J=0,1$ diquarks respectively, are $G(2)$ specific. A tentative spectrum for the bosonic diquarks from lattice simulations has been proposed in \citep{masses2014}. $G(2)$ and $SU(3)$ also share glueballs states, for any numbers of $G(2)$ gluons (2 and 3 in the ground states) and hexaquarks\footnote{The complete explicit decompositions of the products are:\\
\scriptsize{$\{7\} \otimes \{14\} \otimes \{14\} \otimes \{14\} = \{1\} \oplus 10 \{7\} \oplus 6 \{14\} \oplus 15 \{27\} \oplus 20 \{64\} \oplus 13 \{77\} \oplus 13 \{77'\}\oplus 10 \{182\}\oplus 15 \{189\}\oplus 9 \{286\}\oplus 3 \{378\}\oplus 6 \{448\}\oplus 3 \{729\}\oplus \{896\}\oplus 2 \{924\}$

$\{7\} \otimes \{7\} \otimes \{7\} \otimes \{7\} = 4\{1\} \oplus 10 \{7\} \oplus 9 \{14\} \oplus 12 \{27\} \oplus 8 \{64\} \oplus 6 \{77\} \oplus 2 \{77'\}\oplus \{182\}\oplus 3 \{189\}$

$\{7\} \otimes \{7\} \otimes \{7\} \otimes \{7\}\otimes \{7\} = 10\{1\} \oplus 35 \{7\} \oplus 30 \{14\} \oplus 45 \{27\} \oplus 40 \{64\} \oplus 30 \{77\} \oplus 11 \{77'\}\oplus 10 \{182\}\oplus 20 \{189\}\oplus 5 \{286\}\oplus \{378\}\oplus 4 \{448\}$

$\{7\} \otimes \{7\} \otimes \{7\} \otimes \{7\}\otimes \{7\}\otimes \{7\} = 35\{1\} \oplus 120 \{7\} \oplus 120 \{14\} \oplus 180 \{27\} \oplus 176 \{64\} \oplus 145 \{77\} \oplus 65 \{77'\}\oplus 65 \{182\}\oplus 120 \{189\}\oplus 5 \{273\}\oplus 40 \{286\}\oplus 15 \{378\}\oplus 40 \{448\}\oplus \{714\}\oplus 9\{729\}\oplus 5 \{924\}$

$\{14\} \otimes \{14\} = \{1\} \oplus \{14\} \oplus \{27\} \oplus \{77\} \oplus \{77'\}$

$\{14\} \otimes \{14\} \otimes \{14\} = \{1\} \oplus \{7\} \oplus 5 \{14\} \oplus 3 \{27\} \oplus 2 \{64\} \oplus 4 \{77\} \oplus 3 \{77'\}\oplus\{182\}\oplus 3 \{189\}\oplus \{273\}\oplus 2\{448\}$}
}.
Even collective manifestations of $G(2)$ exceptional matter could be different: for example, a $G(2)$-QCD neutron star could display a distinct behavior with respect to a $SU(3)$ neutron gas star, as discussed in \citep{Hajizadeh_2017}. \\
In the next section the focus will be on the phenomenology of the massive $G(2)$ glueball states.

\section{The exceptional gluonic content of the theory: a possible dark matter phenomenology}\label{darkmatter}

As discussed before, the $G(2)$ extension of the SM produces an \textit{exceptional} particle sector: if we move away the six $G(2)$ gluons from the dynamics, these bosons must be secluded and separated from the visible SM sector in first approximation, without experimentally accessible electroweak interactions, unlike WIMPs, and extreme energies (and distances) should be mandatory to access the $G(2)$ string breaking. This could be due to the very high energy scale of the $G(2)-SU(3)$ phase transition, occurring at much greater energies than electroweak breaking scale. This could be the realization of a \textit{beyond Naturalness} criterion. Indeed, $G(2)$ gluons, as $SU(3)$ ones, are electrically neutral and immune to interactions with light and weak $W$, $Z$ bosons at tree level. Another advantage of a $G(2)$ broken theory is that no additional families are added to the Standard Model, unlike $SU(N)$ theories. \\ 
Overall, this seems to be a good scenario for a cold dark matter (CDM) theory (cold means non-relativistic and refers to the standard Lambda-CDM cosmological model), if we find a stable or long-lived candidate. Many vector bosons composite states have been proposed as DM candidates in the last two decades: light hidden glueballs \citep{Juknevich_2009,Juknevich:2009gg,Boddy_2014,Yamanaka:2014pva}, gluon condensates \citep{Klinkhamer_2010}, exceptional dark matter referring to a composite Higgs model with $SO(7)$ symmetry broken to the exceptional $G(2)$ \citep{Ballesteros_2017}, $SU(N)$ vector gauge bosons \citep{Koorambas:2013una}, vector Bose-Einstein condensates (BECs) \citep{Yukawa_2012} and, in general, non-Abelian dark forces \citep{PhysRevD.97.075029}. These studies demonstrate the growing interest in beyond SM non-Abelian frameworks where to develop consistent DM theories, without invoking string theory and keeping the theoretical apparatus sufficiently minimal.\\
In our case, the six \textit{dark} gluons can form dark glueballs constituted by two or three (or multiples) $G(2)$ gluons, according to $\{14\}\otimes \{14\} = \{1\} \oplus ...$ and $\{14\}\otimes \{14\}\otimes \{14\} = \{1\} \oplus ...$ representations \citep{masses2014}, with integer total angular momentum $J = 0, 2$ and $J = 1, 3$ for 2-gluons and 3-gluons balls respectively. In principle, exceptional-colored broken-$G(2)$ glueballs should not be stable, if these massive composite states themselves have no extra symmetries to prevent their decay; since the proposed $G(2)$ theory includes QCD, unlike hidden Yang-Mills theories with no direct connections with the SM \citep{SONI2017379,Juknevich:2009gg}, there exist states that couple to both the exceptional-colored glueballs and $SU(3)$ particles (for example the $G(2)$-breaking Higgs field): hence, whether at tree-level or via loops, these heavy glueballs would not be stable. To avoid this, first of all the new $G(2)$ Higgs should be at least more massive than the lightest 2-gluons glueball, so that $M_H>M_{GG}$, which implies the qualitative constraint $\sqrt{2\lambda} > 2 g_G$ from Eq.(\ref{mH}), (\ref{mGG}); secondly, the decays into meson states should be also forbidden.  In principle, a lightest $J^{PC}=0^{++}$ state, in analogy with standard QCD, could dominate the glueball spectrum \citep{PhysRevLett.77.2622}, but this could be unstable, like the lightest meson $\pi^0$ and the other known scalar particle, the SM Higgs $h^0$. The possibility of a conserved charge or a peculiar phenomenon which guarantees stability to the lightest glueball states should not be ruled out \textit{a priori}, considering that analytical and topological properties of Yang-Mills theory solutions are still not completely understood: even the fundamental problem of color confinement has not a definitive answer nor an analytical proof.\\ 
For example, in analogy with the baryon number conservation and the forbidden proton decay into $\pi^0$, one can introduce a conserved additive \textit{gluon number} $\Gamma$ for the glueball states, which counts the number of massive $G(2)$ gluons (and ``antigluons"), preventing the glueball from decaying into SM mesons, which are not made of $G(2)$ gluons (one has to keep in mind that these peculiar bosons do mantain the color quarks/antiquarks quantum number). Indeed, also the $U(1)_B$ global symmetry of the Standard Model which prevents the proton decay is an \textit{accidental symmetry} and not a fundamental law, that can be broken by quantum effects. A large class of models \citep{2015PhLB..746..430A,Bernal_2016,Branco_2012} imposes global discrete $Z_2$ (or generally $Z_n$ \citep{Yaguna_2020}) or continuous $U(1)$ symmetries to guarantee DM stability, in which DM is odd under the new symmetry while SM fields are assumed to be even: even in our case one can introduce such a $Z_2$ symmetry, inherited from the new $G(2)$--breaking Higgs field, preventing the lightest $G(2)$ two--gluons glueball decay. A complementary choice is to invoke another multiplicative quantum number, \textit{i.e.} a $G$--parity conservation for a generic Yang--Mills theory as suggested in \citep{Bai_2010}, to generalize the $C$--parity and apply it to meson-like multiplets; unlike the lightest QCD mesons $\pi$ which possess electroweak interactions, the lightest $G(2)$ glueball could represent a sort of stable dark meson (due to the quark--antiquark quantum numbers carried by the two $G(2)$ gluons), whose dynamics is constrained inside the $G(2)$ broken sector itself, leading to an exactly conserved $G$--parity and the impossibility to decay into $G$--even SM particles. 
\\
We have the same lack of knowledge for the glueballs interactions with their own environments: as for residual nuclear force between hadrons in nuclei, the possibility of a residual binding interaction between glueballs, preventing the decay (like the neutron case in the nucleus), must be investigated.\\ 
Possibly, one can also postulate a suppression scale $1/f_G(\Delta M)$ for the couplings with SM which depends on the relative mass difference between the interacting particles, \textit{i.e.} the dark glueball and the quarks: if the masses of the glueballs are too high w.r.t. the QCD scale, their decay might be highly suppressed. \\
Another interesting opportunity is to invoke the $J^{PC}=0^{++},2^{++}$ dark $G(2)$ glueball states as graviton counterparts in a AdS/CFT correspondence framework, as discussed in \citep{Rinaldi_2018} for QCD glueballs (even if this does not exhaust the quest for stability). Moreover, the scalar glueball could be part of a scalar-tensor gravity approach \citep{scalarvectorgravity}, whereas the tensor glueball could play the role of a massive graviton-like particle, for example in the context of bimetric gravity theory \citep{Akrami_2015,PhysRevD.94.084055}, where the massive gravitational dark matter can non-trivially interact with gravity itself. The couplings to SM quarks of such a tensor DM can be by far too weak \citep{PhysRevD.94.084055}, making it undetectable in collider searches; besides, the requirement of a correct DM abundance and stability constrains a non-thermal spin-2 mass to be $>> 1$ TeV \citep{Babichev_2016}.\\
Furthermore, it must be considered that bosonic ensembles could eventually clump together to form a BEC: once the temperature of a cosmological boson gas is less than the critical temperature, a Bose-Einstein condensation process can always take place during the cosmic history of the Universe, even if the not low mass of these candidates should disfavor this scenario. For example, the occurence of glueballs condensates and glueball stars have been recently discussed in \citep{SONI2017379,Giacosa:2017eqy,Lucini:2014paa,daRocha:2017cxu}.\\

Such a dark sector can naturally accommodate the fact that there is only gravitational evidence for dark matter so far, certainly disfavoring direct and indirect searches, and it can also qualitatively account for the observation that dark matter and ordinary matter are in commensurable quantities (approximately 5:1 from recent Planck experiment measurements \cite{Aghanim:2018eyx}), as they come from the same broken gauge group. Given the forbidden or extremely weak interactions between the $G(2)$ glueball states and ordinary matter, the usual WIMP-like scenario in which the DM relic abundance is built via the freeze-out mechanism cannot be achieved, since these bosons are not in thermal equilibrium with the baryon-photon fluid in the early Universe: their production should be abruptly triggered by a first-order cosmological phase transition, possibly fixing their initial abundance. It is well-known that several non freeze-out models has been proposed in literature, such as the \textit{FIMP} (Feebly Interacting Massive Particle) cosmology via a freeze-in mechanism \citep{Bernal_2017,Hall_2010}, in which the comoving DM abundance freezes when the number densities of the visible sector, generating DM by decays or annihilations, become Boltzmann-suppressed, ending the yield. This requires an extremely small coupling ($O(10^{-7})$ or less) with the visible sector. Another intriguing alternative is represented by the so-called \textit{Dark freeze-out}, for which DM reaches an equilibrium heat bath within the dark sector itself, never interacting with SM particles: in this case, the dark ensemble was initially populated by a freeze-in-type yield from part of the visible sector.\\
In analogy, trying to construct a particle cosmology for the exceptional gluonic content of the present theory, let's suppose that a two-gluons $G(2)$ glueball could be \textit{ab initio} produced out of thermal equilibrium, for instance by a heavy mediator decay in the visible sector heat bath, such as the new $G(2)$--Higgs ($H\rightarrow DM DM$): this is viable if $M_H \geq M_{GG}$ and the coupling between DM and the heavy Higgs is sufficiently weak, realizing an \textit{exotic-Higgs portal DM} \citep{Bernal_2019}.\\ 
In principle, the most general scalar glueball effective potential, in the large $N$ limit of a $SU(N)$ gauge theory, may contain not only a quartic interaction, proper of a Higgs portal, but also the cubic and higher order terms \citep{SONI2017379,Yamanaka:2019yek,Soni:2016gzf}, in the form
\begin{equation}\label{effpot}
V(S) = \sum_{i=2}^{\infty}{\frac{a_{i}}{i!}{\Big(\frac{4\pi}{N}\Big)}^{i-2} m^{4-i}S^i}
\end{equation}
where $S$ is the scalar $G(2)$ glueball field, $m$ the mass term and the coefficients are $a_i\approx 1$, which could be obtained from lattice computations. The trilinear interaction might generate an attractive Yukawa-like potential \citep{Yamanaka:2019yek}, whereas the quartic one may be repulsive, according to the sign of the coupling; the fifth and higher terms might be suppressed by the mass scale and the decreasing couplings.
Choosing the simplest scalar case with a $Z_2$ symmetry with a negligible cubic interaction (according to the minimal Higgs-portal paradigm), from now on we adopt a renormalizable scalar $G(2)$-glueball potential, coupled to the exotic heavy Higgs sector, of the type
\begin{equation}\label{cosmology}
V(\Phi,S) = V(\Phi) + \frac{m^2}{2}S^2+\frac{\lambda_S}{4} S^4+\frac{\lambda_{HS}}{2} \Phi^2 S^2
\end{equation}
where $\lambda_S$ is the quartic self-interaction strength, $\lambda_{HS}$ the heavy Higgs-scalar glueball coupling and $V(\Phi)$ is described by Eq.~\eqref{Hpot}; ${M_S}^2=m^2+\lambda_{HS} w^2 /2$ can be defined as the total mass after the $G(2)$ symmetry breaking (which trivially implies $\lambda_{HS} \leq 2 {M_S}^2 / w^2$). If the portal coupling is sufficiently small \citep{Bernal_2017}, one can recover a correct dark matter relic abundance $\Omega h^2 \simeq 0.12$ \cite{Aghanim:2018eyx}. In fact, the approximate solution for this present-day DM abundance, assuming the initial number density of DM particles negligible, is \citep{Bernal_2017}
\begin{equation}\label{freezein}
\Omega_{S}h^2 \simeq 4.48\times 10^{8}\frac{g_H}{g_{*s}\sqrt{g_*}} \frac{M_{S}}{\rm GeV} \frac{M_{Pl}\,\Gamma_{H\rightarrow SS}}{M_{H}^2}\,,
\end{equation}
where $g_{*s}$ and $g_*$ are effective numbers of degrees of freedom for entropy and energy densities, $g_H$ is the intrinsic number of degrees of freedom of $H$ (the expression is evaluated around $T\simeq M_{H}$) and where $M_{Pl}$ is the Planck mass. For $g_{*s}\simeq g_*$ and $\Gamma_{H\rightarrow SS}\simeq {{\lambda^2}_{HS}}\,M_H/(8\pi)$, Eq.~\eqref{freezein} gives an estimate for the coupling intensity as a function of the DM abundance:
\begin{equation}\label{freezein2}
\lambda_{HS}\simeq 10^{-12}\Big(\frac{\Omega_S h^2}{0.12}\Big)^{1/2}\Big(\frac{g_*}{100}\Big)^{3/4}\Big(\frac{M_H}{M_S}\Big)^{1/2}.
\end{equation}
It's easy to verify that, for $\Omega h^2 \simeq 0.12$, $g_*\simeq 100$ and not $M_H/M_S >> 1$ ratios, $\lambda_{HS}<<1$. Furthermore, studying the exotic Higgs-portal \citep{Bernal_2017} in the context of the $G(2)$-breaking mechanism, one can approximate again Eq.~\eqref{freezein}--\eqref{freezein2} as $\Omega_{S}h^2/0.12 \simeq 10^{24}{{\lambda_{HS}}^2}\frac{M_{S}}{M_{H}}$ and insert the scalar glueball mass for $m^2 << \lambda_{HS} w^2$, \textit{i.e.} a scalar mass integrally generated by the Higgs expectation value
\begin{equation}\label{gmassint}
M_S=M_{GG}\simeq 2 g_G w \approx \sqrt{\lambda_{HS}/2}w, 
\end{equation}
as an explicit function of the Higgs-scalar coupling, to obtain:
\begin{equation}
\Omega_{S}h^2/0.12 \simeq 10^{24}{{\lambda_{HS}}^{5/2}} (w/M_H).
\end{equation}
For $w/M_H=O(1)$, to reproduce the correct relic density the coupling should be very tiny, \textit{i.e.} $\lambda_{HS}\sim 10^{-(10\div9)}$, as in most FIMP theories \citep{Bernal_2017,Enqvist_2018}.\\
The situation can change if $\lambda_S$ is large enough ($\lambda_S > 10^{-3}$ \citep{Bernal_2016}), \textit{i.e.} if scalar self-interactions are active: the DM particles, initially produced by the previous mechanism, may thermalize among themselves even if the dark sector consists of only one particle species, due to number-changing processes (generic $n_{DM}\rightarrow n'_{DM}$ processes), which reduce the average temperature of DM particles and increase the number density until equilibrium is reached. The resulting relic abundance could therefore change even though the coupling between the visible and dark sector is absent. This could be the case of a \textit{SIMP} (Strongly Interacting Massive Particle) scenario with a dark freeze-out mechanism \citep{Bernal_2017,Bernal_2016, Bernal_2019}. For example, in case of a quartic self-interaction, thermalization of the dark sector within itself through $2 \leftrightarrow 4$ scatterings is active if $\lambda_S$ exceeds a critical value, so that DM reaches a thermal equilibrium. When DM is no longer relativistic, $4\rightarrow2$ processes dominate the dynamics in the so-called \textit{cannibalization era}, which ends when its rate drops below the Hubble rate, fixing the DM number density to a modified yield through the dark freeze-out. Odd processes, such as $2 \leftrightarrow 3$, could be neglected for DM potential without odd powers terms and are forbidden for multiplicative $Z_n$ symmetry conservation for even $n$ \citep{Bernal_2016}, like in most well-established DM setups \citep{Bernal_2016bis}. As shown in \citep{Bernal_2016,Heikinheimo_2017}, if DM relic abundance is solely computed via a Higgs decay in a freeze-in framework, without \textit{dark thermalization} for $4\rightarrow2$ interactions, it could be appreciably underestimated (over an order of magnitude) and, consequently, the previous bounds for the parameters space could be altered. 
Quartic self-interactions should also fulfill additional constraints from cosmology \citep{Heikinheimo_2017,Enqvist_2018}, such as the isocurvature bound, for which the scalar mass is bounded from above:
\begin{equation}\label{inflat}
\frac{M_S}{\text{ GeV}} < 6 {\lambda_S}^{3/8}\Big(\frac{H_*}{10^{11} \text{ GeV}}\Big)^{-3/2}
\end{equation}
with $H_*$ the Hubble parameter at the inflationary scale.\\
The resolution of Boltzmann equation for SIMP DM usually leads to scenarios where the dark freeze-out temperature is less than the visible ensemble one, making DM naturally colder than SM particles. See \citep{Bernal_2016bis, Choi_2017,Forestell:2016qhc,Bhattacharya_2020,Choi:2020ara, 2017arXiv170401804S} for insights regarding $3\rightarrow2$ dark thermalization and explicit formulas for the relic density for a SIMP DM with $3 \leftrightarrow 2$ annihilations, which do not have immediate analytic representations.\\
For smaller interactions with the visible sector, the thermal production of DM particles is insignificant and DM must come from a non-thermal mechanism, leading to a Super-WIMP (SWIMP)-like scenario, for example through a direct DM-producing inflaton decay \citep{Allahverdi:2002nb,Almeida:2018oid,delaMacorra:2012sb}, if the heavy Higgs scalar responsible for the $G(2)\rightarrow SU(3)$ transition is identified as the inflaton field. In this \textit{inflaton-portal} case \citep{Heurtier:2019eou}, the glueball abundance is basically fixed by few parameters, \textit{i.e.} the reheating temperature $T_{rh}$, the inflaton mass and branching ratio into $G(2)$ gluons/scalar glueball $B_S$ \citep{Allahverdi:2002nb}:\\
\begin{equation}
\Omega_{S}h^2 \simeq 2 \times 10^{8} B_S (M_S/M_H) \frac{T_{rh}}{\rm GeV}
\end{equation}
For a cosmological reheating above the GeV scale, the $B_S (M_S/M_H)$ factor should be quite small, \textit{i.e.} $< 10^{-9}$. 
The scenario works well especially for large DM masses, between the weak scale and the PeV scale \citep{Almeida:2018oid}, and for extremely decoupled EeV candidates \citep{Heurtier:2019eou}. \\

Finally we are going to briefly discuss possible manifestations of $G(2)$ gluons in the present Universe. The six massive gluons from the broken exceptional $G(2)$ group, mainly as stable $J^{PC}=0^{++}$ (and possibly $J^{PC}=2^{++}$) glueballs, could clump and organize into dark matter halos, in form of a heavy bosons gas or in some fluid systems \citep{Mirza_2011,Harko_2011}, for sufficiently low temperatures and/or enough high densities. Indeed, the idea of a (super)fluid dark matter has recently attracted attention in literature, from Khoury's promising proposal of a unified superfluid dark sector \citep{Sharma:2018ydn,Ferreira_2019}. The fact that dark matter particles could assume different ``phases'' according to the environment (gas, fluid or BECs) is very suitable to account for the plethora of dark matter observations at all scales, from galactic to cosmological ones. The realization of this scenario usually involves axions or ALPS (axion like particles) \citep{Axion}, which are light or ultralight (with masses from 10${}^{-24}$ to 1 $\mathrm{eV}$ in natural units) and capable of reproducing DM halos properties: dark matter condensation and self-gravitating Bose-Einstein condensates have been extensively studied in \citep{Book_BEC,B_hmer_2007,Chavanis:2016ial}.\\
In our scenario we hypothesize the main constituents of a possible dark fluid are broken-$G(2)$ massive gluons composite states. They are certainly massive, so that we cannot invoke the axion-like description. But still $G(2)$ glueballs could aggregate into extended objects: one can explore the possibility that heavy $G(2)$ gluon dark matter is capable of producing stellar objects, which could populate the dark halos. Many models of exotic stars made of unknown particles have been proposed, especially for sub-GeV masses, such as bosonic stars (where the particle is a scalar or pseudoscalar \citep{Sanchis-Gual:2019ljs,Eby:2015hsq,Liebling_2017}, most likely for a quartic order repulsive self-interaction \citep{Liebling_2017}), Proca stars (for massive spin 1 bosons \citep{Brito:2015pxa,Landea_2016,Minamitsuji_2018} -- one could invoke a correspondence with 3-gluons glueballs), BEC stars \citep{Chavanis_2012}, or QCD glueballs stars \citep{daRocha:2017cxu}. The key ingredients to try to build up stellar objects with non--light bosons are mainly the magnitude of the quartic order self-interaction $\lambda_4$ of the constitutive boson and its mass. Our glueball scalar field $S$ evolving in the General Relativity framework can be described by the Einstein-Klein-Gordon (EKG) action,
\begin{equation}
	\label{EKG}
	\mathcal{S} = \int \mathrm{d}^4x\sqrt{-g} \left[\frac{R}{16\pi G} - g^{\mu\nu}\partial^\mu\bar S\partial_\nu S - V(S)\right]\,,
\end{equation}
where $V(S)$ is the bosonic potential, $\bar S$ the complex conjugate for a complex scalar, $R$ is the Ricci scalar, $g^{\mu\nu}$ is the metric of the space--time, $g$ its determinant and $G$ is Newton's gravitational costant. The variation of the action with respect to the metric leads to the related Einstein equations \citep{Liebling_2017,Cardoso_2019} for static, spherically--symmetric geometries. To build up a massive boson star, a quartic self-interaction potential is needed to balance the inward gravitational force, in the form of the previously described (Eq.\ref{cosmology}) potential, \textit{i.e.} $V({\left|S\right|}^2)=\frac{M^2_{S}}{2}{\left|S\right|}^2+\frac{\lambda_4}{4}{\left|S\right|}^4$, where $\lambda_S\equiv\lambda_4>0$ ($\lambda_4<0$) signifies a repulsive (attractive) interaction: repulsive self-interactions can give rise to very dense boson stars. The maximum mass for such an object is \citep{Chavanis_2012,Hertzberg_2021,Liebling_2017}
\begin{equation}
M_{max}\sim (0.1\div 1){\sqrt{\lambda_4}\,M_{Pl}^3\over M^2_{S}/{\rm GeV}^2}
\label{Maxmass}
\end{equation}
where $M_{Pl}\equiv1/\sqrt{G}\approx 1.2\times 10^{19} GeV \sim M_{\odot}^{1/3}$ is the Planck mass and $M_{\odot}$ the solar mass, with a consequent boson star radius \citep{SONI2017379}
\begin{equation}
R_{BS}\sim (0.1\div 1){\sqrt{\lambda_4}\,\times 10\, {\rm km} \over M^2_{S}/{\rm GeV}^2}.
\end{equation}
For example, for a heavy $M_S\sim 10$ TeV scalar, the maximum values, as a function of the self-interaction, are $M_{max}\lesssim 10^{-8}\sqrt{\lambda_4} M_{\odot}$ and $R_{BS}\lesssim 10^{-4}\sqrt{\lambda_4}\,$ m: the resulting object carries a small mass w.r.t. usual stars (about one hundredth of the Earth mass) and a sub--millimeter radius for not two high $\lambda_4$, making it a tiny ball of glueballs. Instead, for a 1 GeV scalar, $M_{max}\lesssim \sqrt{\lambda_4} M_{\odot}$ and $R_{BS}\lesssim \sqrt{\lambda_4}\,\times 10\, {\rm km}$ stand, producing a solar mass object with a neutron star--like radius. 
As stressed in \citep{Hertzberg_2021}, at the maximum mass the radius is slightly larger than the corresponding Schwarzschild radius. In addition, for $\lambda_4 = O(1)$, the relation is similar to the Chandrasekhar mass for white dwarf stars $M_{wd}\approx \frac{M_{Pl}^3}{m^2_{p}}$: this could resemble the case of $SU(N)$ glueballs, estimating the quartic coupling as $\lambda_4\sim (4\pi)^2/N^2$ \citep{Hertzberg_2021} (see Eq.~\eqref{effpot}), being $\lambda_4 < 4\pi$ for perturbativity \citep{Eby:2015hsq}. Even if it is fair that $G(2)$ glueballs are strongly self interacting, it is quite hard to make precise estimates for their scattering process, given our general limited knowledge concerning strongly coupled theories. Nevertheless, considering the dimension of the exceptional $G(2)$ group and its lattice behaviour \citep{thermoG22014}, we could qualitatively approximate it to a $SU(4)$ theory and obtain $\lambda_4\sim \pi^2$, which almost saturates the perturbativity condition $\lambda_4 < 4\pi$. For example, for such a $\lambda_4$, both the aforementioned $M_{max}$ ranges, \textit{i.e.} $10^{-8}M_{\odot}$ and few $M_{\odot}$, fall into mass windows with weak constraints for MACHOs from up-to-date microlensing analysis \citep{Calcino:2018mwh,Brandt:2016aco}.
Obviously, the feasibility of a star made of self-interacting scalars also depends on a correct estimate of the possible $3\rightarrow1$, $3\rightarrow2$ and $4\rightarrow2$ annihilation processes inside the star \citep{Hertzberg_2021}, which in turn depend on the symmetries (\textit{e.g.} $Z_n$) of the exotic sector; for a non-conserved glueballs number, \textit{i.e.} for a real scalar with no additional symmetries, $3/4 \rightarrow 2$ processes could trigger the decay of massive bosons stars, leading to severe bounds on their mass: $M_{max}<10^{-11}M_{\odot}$.\\
For the so--called mini boson stars (with no scalar interaction potential) and Proca stars, the maximum mass is quite smaller and scales with $\sim \frac{M_{Pl}^2}{M_{S}}$. Furthermore, unlike complex scalar fields, real scalars stars could not possess the required stability: however, with a non-trivial time-dependent stress-energy tensor, long-term stable oscillating geometries can be achieved \citep{Cardoso_2019}. These ``oscillatons" mass-radius relations are indeed very similar to the boson stars ones. \\
It is well-known that the inclusion of self-interactions in the DM sector should be in agreement with the upper limits which come from several astrophysical observations \citep{Bertone,Harvey_2019,Bernal_2020}, mainly from colliding galaxy clusters dynamics, like the the Bullet Cluster \citep{Robertson_2016}. 
The DM-DM scattering cross section must qualitatively satisfy $\frac{\sigma}{M_S} < 1 \frac{\text{cm}^2}{\text{g}}\sim 10^{-24}\frac{\text{cm}^2}{\text{GeV}}$. Assuming $\sigma = \frac{9{\lambda_4}^2}{32\pi M^2_{S}}$ at tree level \citep{Bernal_2017}, the bound \citep{Eby:2015hsq} results in 
\begin{equation}\label{bounds}
 |\lambda_4| < 4 \cdot 10^{2}\Big(\frac{M_S}{\text{ GeV}}\Big)^{3/2}
\end{equation}
which is consistent with the one obtained in \citep{Bernal_2019}. In principle the intensity of the self-interaction is not effectively bounded by the scalar mass scale of the present theory: the more the DM particle is massive the more the constraint for $\lambda_4$ is relaxed. Exploiting Eq.~\eqref{gmassint} with a FIMP-like $\lambda_{HS}\sim 10^{-10}$, Eq.~\eqref{bounds} becomes
\begin{equation}
|\lambda_4| < 10^{-5}\Big(\frac{w}{\text{ GeV}}\Big)^{3/2}.
\end{equation}
For the aforementioned $\lambda_4\sim \pi^2$ self-interaction assumption, this formula implies an expectation value for the heavy Higgs satisfying $w \geq 10^4$ GeV, which is completely consistent with a beyond SM phenomenology. Such a FIMP/SIMP-like singlet scalar model, constrained by astrophysical observations and the perturbativity condition, naturally leads to a DM mass in the GeV/sub-GeV region \citep{Bernal_2016} (Eq.~\eqref{bounds}). This is also valid from Eq.~\eqref{inflat} for an inflation scale $H_*\gtrsim 10^{12}$ GeV. The empirical Eq.~\eqref{bounds} relation for the glueball scalar mass is quite equivalent to the theoretical ones in Eq.~\eqref{freezein2}--\eqref{gmassint}, for which $\lambda_{HS}\sim 10^{-10}$ and $w/M_H=O(1)$ imply $M_S\sim 10^{-(5\div 4)} M_H$. \\
DM candidates could be certainly studied from peculiar behaviors of the compact astrophysical objects they form, characterizing physical observables useful to disentangle standard scenarios from exotic phenomenology \citep{Cardoso_2019}. In fact, for a few years we can take advantage of both electromagnetic and gravitational waves astronomy as powerful probes to discriminate compact objects as a function of their ``compactness" (or ``closeness" to a black hole -- a quantity related to the mass to radius ratio $M/R$ is a possibility), ``shadow" and gravitational waves emission \citep{Cardoso_2019}. For example, a hypothetical binary $SU(N)$ gluons star could be disentangled from a binary black hole system, due to possible differences in the gravitational wave frequency and amplitude, as demonstrated in \citep{SONI2017379}: this comes from the fact that the mass-radius relation of the glueball dark star, as in our case, is different from that of a black hole, as discussed before. In addition, in \citep{SONI2017379} the authors show that, adopting a large $N$ glueball potential like the one in Eq.~\eqref{effpot}, the dark star is allowed to be more massive and larger at the same time, w.r.t. the quartic potential case in Eq.~\eqref{cosmology}. No accompanying luminous signal is obviously expected for $G(2)$ glueballs astrophysical objects, unlike generic beyond SM theories equipped with electro-weak interactions.\\
Even more ambitious, one should consider the possibility to probe the $SU(3)-G(2)$ phase transition from black hole formation, as suggested in \citep{Ohnishi:2011jv} for quark matter: this is possible if black holes are formed through exotic condensed matter stages beyond degenerate neutron matter, maybe exploiting the theoretical framework of gauge/gravity dualities \citep{Critelli:2017oub}. It could be also worthwhile to speculate on the possibility that, in extremely high pressure and temperature quark matter phases, unbroken $G(2)$ quarks can combine into multiquarks particles or can be seized by $G(2)$ gluons to form $qGGG$ screened states, rearranging QCD matter into a phase of color-singlet hybrids and exotic hadrons. For example, the preliminary work by \citep{Hajizadeh_2017} demonstrates the possibility to distinguish a simplified $G(2)$--QCD neutron star using the mass-radius relation.\\
So dark $G(2)$ glueballs can be very versatile and exploitable within theoretical speculations, especially for exotic scalars coupled to a Higgs sector. 

\section{Conclusions}

If Nature physical description is intrinsically mathematical, fundamental microscopic forces might be manifestation of the algebras that can be built via the Cayley-Dickson construction process. In other words, algebras can guide physics through the understanding of fundamental interactions. To translate the mathematical meaning into a physical language, one has to move from abstract algebras to groups of symmetry, through a correspondence here proposed as an \textit{automorphism relation}. This leads to the discovery of a mismatch between $SU(3)$ strong force and octonions: the octonions automorphism group is the exceptional group $G(2)$, which contains $SU(3)$, but it is not exhausted by $SU(3)$ itself. In the difference between the physical content of $G(2)$ and $SU(3)$ new fundamental particles lie, in the form of six additional massive bosons organized in composite states, disconnected by Standard Model dynamics: \textit{the exceptional-colored $G(2)$ gluons}. These gluons cannot interact with SM particles, at least at the explored energy scales, due to their mass and QCD string behavior. \textit{Mathematical realism} has been the guide and criterion to build this minimal extension of the Standard Model.\\  
Hence, for the first time in literature, $G(2)$ was treated and developed as a realistic symmetry to enlarge the Standard Model, and not only as a lattice QCD tool for computation. $G(2)$ is a good gauge group to describe a larger interaction, which operated in the Early Universe before the emergence of visible matter; when the Universe cooled down, reaching a proper far beyond TeV energy scale at which $G(2)$ is broken, usual $SU(3)$ QCD appeared, while an extra Higgs mechanism produced a secluded sector of cold exceptional-colored bosons. The SM is naturally embedded in this framework with a minimal additional particle content, \textit{i.e.} a heavy scalar Higgs particle, responsible for a Higgs mechanism for the strong sector symmetrical w.r.t. the electroweak one, and a bunch of massive gluons, in principle with the same masses, whose composite states might play the role of dark matter. Some \textit{accidental} stability mechanisms for the dark glueballs have been proposed. The presence of a new Higgs field represents the usual need of a scalar sector to induce the symmetry breaking of a fundamental gauge symmetry. The extra Higgs might have visible decay channels, but it belongs to a very high energy scale which is certainly beyond current LHC searches, \textit{i.e.} a multi-TeV or tens/hundreds of TeV scale defined by the vaccum expectation value of the extra Higgs. The resulting exceptional glueball DM is certainly compatible with direct, indirect, collider searches and astrophysical observations, as it is almost collisionless. Several cosmological scenarios for these candidates were discussed, constraining the parameters space of the theory.\\
In addition, if one tries to extend the correspondence between mathematical algebras and physical symmetries further beyond octonions, sedenions show an intriguing property: they still have $G(2)$ as a fundamental automorphism, but ``tripled" by an $S_{3}$ factor, which resembles the three fermion families of the Standard Model and its $S_3$-invariant extension. We know larger symmetries can be constructed using the products of octonions (sedenions) and the other division algebras, pointing towards subsequent exceptional groups, as illustrated by the Freudenthal--Tits magic square, which have been the subjects of string theories, but we did not want to push the dissertation in this direction. Indeed the choice of $G(2)$, as automorphism group of octonions and minimally enlarged non-Abelian compact Lie algebra of rank 2, is the minimal exceptional extension of the Standard Model including a reliable exotic sector, requiring no additional particle families nor extra fundamental forces. This fact reconciles the particle \textit{desert} observed between the SM Higgs mass scale and the TeV scale. $G(2)$ could guarantee peculiar manifestations in extreme astrophysical compact objects, such as boson stars made of $G(2)$ glueballs, which can populate the dark halos and be observed in the future studying their gravitational waves and dynamics.\\
The development of a definitive theory is beyond the purpose of the present phenomenological proposal, which is intended as a guideline for further speculations. 


\bibliography{bibliography} 

\begin{thebibliography}{100}

\bibitem{ModernPP}
M.~Thomson, {\em Modern particle physics}.
\newblock Cambridge University Press, 2013.

\bibitem{Bertone}
G.~Bertone, {\em Particle Dark Matter: Observations, Models and Searches}.
\newblock Cambridge University Press, 2010.

\bibitem{ProfumoDM}
S.~Profumo, {\em An introduction to particle dark matter}.
\newblock World Scientific Publishing Europe Ltd, 2017.

\bibitem{InfraredG}
E.~Mitsou, {\em Infrared non-local modifications of general relativity}.
\newblock Springer, 2016.

\bibitem{ModG_largeDist}
P.~Eleftherios, {\em Modifications of Einstein's theory of gravity at large
  distances}.
\newblock Springer, 2015.

\bibitem{Clifton:2011jh}
T.~Clifton, P.~G. Ferreira, A.~Padilla, and C.~Skordis, ``{Modified Gravity and
  Cosmology},'' {\em Phys. Rept.}, vol.~513, pp.~1--189, 2012.

\bibitem{baudis_2018}
L.~Baudis, ``The search for dark matter,'' {\em European Review}, vol.~26,
  no.~1, p.~70–81, 2018.

\bibitem{DMATLAS}
S.~Schramm, {\em Searching for dark matter with the ATLAS detector}.
\newblock Springer, 1st~ed., 2017.

\bibitem{DMTevish}
N.~Masi, ``Dark matter: Tev-ish rather than miraculous, collisionless rather
  than dark,'' {\em The European Physical Journal Plus}, vol.~130, 04 2015.

\bibitem{Salvio:2019agg}
A.~Salvio and F.~Sannino, eds., {\em {From the Fermi Scale to Cosmology}}.
\newblock Frontiers, 2019.

\bibitem{giudice2017dawn}
G.~F. Giudice, ``{The Dawn of the Post-Naturalness Era},'' in {\em From My Vast
  Repertoire ...: Guido Altarelli's Legacy} (A.~Levy, S.~Forte, and G.~Ridolfi,
  eds.), pp.~267--292, 2019.

\bibitem{Young:2016ala}
B.-L. Young, ``{A survey of dark matter and related topics in cosmology},''
  {\em Front. Phys.(Beijing)}, vol.~12, no.~2, p.~121201, 2017.

\bibitem{Hooper:2018kfv}
D.~Hooper, ``{TASI Lectures on Indirect Searches For Dark Matter},'' {\em PoS},
  vol.~TASI2018, p.~010, 2019.

\bibitem{Gaskins:2016cha}
J.~M. Gaskins, ``{A review of indirect searches for particle dark matter},''
  {\em Contemp. Phys.}, vol.~57, no.~4, pp.~496--525, 2016.

\bibitem{NicMario}
N.~Masi and M.~Ballardini, ``A conservative assessment of the current
  constraints on dark matter annihilation from cosmic rays and cmb
  observations,'' {\em International Journal of Modern Physics D}, vol.~26,
  no.~06, p.~1750041, 2017.

\bibitem{Wellegehausen_2011}
B.~H. Wellegehausen, A.~Wipf, and C.~Wozar, ``{Phase diagram of the lattice G2
  Higgs model},'' {\em Physical Review D}, vol.~83, Jun 2011.

\bibitem{Maas:2012ts}
A.~Maas and B.~H. Wellegehausen, ``{$G_2$ gauge theories},'' {\em PoS},
  vol.~LATTICE2012, p.~080, 2012.

\bibitem{Book_Gursey}
F.~Gursey and C.-H. Tze, {\em On the role of division, Jordan, and related
  algebras in particle physics}.
\newblock World Scientific, 1996.

\bibitem{Book_Conway}
J.~H. Conway and D.~A. Smith, {\em On quaternions and octonions: their
  geometry, arithmetic, and symmetry}.
\newblock AK Peters/CRC Press, 2003.

\bibitem{Book_Dixon}
G.~M. Dixon, {\em Division algebras: octonions, quaternions, complex numbers,
  and the algebraic design of physics}.
\newblock Dordrecht; Boston: Kluwer Academic Publishers, 1994.

\bibitem{Clifford_Lounesto}
P.~Lounesto and L.~M. Society, {\em Clifford algebras and spinors}.
\newblock Cambridge University Press, 2nd~ed., 2001.

\bibitem{GRESNIGT2018212}
N.~G. Gresnigt, ``Braids, normed division algebras, and standard model
  symmetries,'' {\em Physics Letters B}, vol.~783, pp.~212 -- 221, 2018.

\bibitem{Book_Maia}
M.~D. Maia, {\em Geometry of the Fundamental Interactions}.
\newblock Springer, 2011.

\bibitem{furey2016standard}
C.~Furey, ``{Standard model physics from an algebra?},'' {\em
  {arXiv:1611.09182}}, 2016.

\bibitem{Manogue_2010}
C.~A. Manogue and T.~Dray, ``{Octonions, E6, and particle physics},'' {\em
  Journal of Physics: Conference Series}, vol.~254, p.~012005, Nov 2010.

\bibitem{gording2019unified}
B.~Gording and A.~Schmidt-May, ``{The Unified Standard Model},'' {\em
  arXiv:1909.05641}, 2019.

\bibitem{Baez:2001dm}
J.~C. Baez, ``{The Octonions},'' {\em Bull. Am. Math. Soc.}, vol.~39,
  pp.~145--205, 2002.

\bibitem{ZeroAnnihil}
D.~C.~I. Daniel K.~Biss, Daniel~Dugger, ``{Large Annihilators in Cayley-Dickson
  Algebras},'' {\em Communications in Algebra}, vol.~36, no.~2, pp.~632--664,
  2008.

\bibitem{Gillard_2019}
A.~B. Gillard and N.~G. Gresnigt, ``Three fermion generations with two unbroken
  gauge symmetries from the complex sedenions,'' {\em The European Physical
  Journal C}, vol.~79, May 2019.

\bibitem{cawagas_2014}
R.~Cawagas, ``{On the structure and zero divisors of the Cayley-Dickson
  sedenion algebra},'' {\em Discussiones Mathematicae. General Algebra and
  Applications}, vol.~24, 01 2004.

\bibitem{Cacciatori:2009qu}
S.~L. Cacciatori and B.~L. Cerchiai, ``{Exceptional groups, symmetric spaces
  and applications},'' {\em arXiv:0906.0121}, 2009.

\bibitem{EVANS198892}
J.~Evans, ``{Supersymmetric Yang-Mills theories and division algebras},'' {\em
  Nuclear Physics B}, vol.~298, no.~1, pp.~92 -- 108, 1988.

\bibitem{Book_Schwerdtfeger}
H.~Schwerdtfeger, {\em Geometry of complex numbers: circle geometry, Moebius
  transformation, non-euclidean geometry}.
\newblock Dover Pubns, 1979.

\bibitem{PhysFSym}
J.~Schwichtenberg, {\em Physics from symmetry}.
\newblock Springer, second~ed., 2018.

\bibitem{Super_Nath}
P.~Nath, {\em Supersymmetry, supergravity, and unification}.
\newblock Cambridge University Press, 2017.

\bibitem{Burgess}
C.~P. Burgess and G.~D. Moore, {\em {The Standard Model: a Primer}}.
\newblock Cambridge University Press, 2012.

\bibitem{Schwinger}
J.~Schwinger, {\em Particles, sources, and fields. Volume 1}.
\newblock Taylor and Francis Ltd, 2018.

\bibitem{Book_Field}
M.~Field, {\em Dynamics and symmetry}.
\newblock London Imperial College Press, 2007.

\bibitem{OpSym_saller}
H.~Saller, {\em Operational symmetries: Basic operations in physics}.
\newblock Springer, 2017.

\bibitem{Aguilar:1998ws}
M.~A. Aguilar and M.~Socolovsky, ``{On the topology of the symmetry group of
  the standard model},'' {\em Int. J. Theor. Phys.}, vol.~38, pp.~2485--2509,
  1999.

\bibitem{Yale}
P.~B. Yale, ``Automorphisms of the complex numbers,'' {\em Mathematics
  Magazine}, vol.~39, no.~3, pp.~135--141, 1966.

\bibitem{Book_Saller}
H.~Saller, {\em Operational quantum theory I: nonrelativistic structures}.
\newblock Springer, 2006.

\bibitem{Bialynicki_Birula_2013}
I.~Bialynicki-Birula and Z.~Bialynicka-Birula, ``{The role of the
  Riemann–Silberstein vector in classical and quantum theories of
  electromagnetism},'' {\em Journal of Physics A: Mathematical and
  Theoretical}, vol.~46, p.~053001, Jan 2013.

\bibitem{Moebius}
R.~A. Gelfand I.~M., Minlos and S.~Z. Ya, {\em Representations of the rotation
  and Lorentz groups and their applications}.
\newblock Oxford Pergamon Press; Dover Publications Inc., 1963, new edition
  2018.

\bibitem{2012IJTP...51.3228P}
T.~L. Pushpa, Bisht~P.S. and O.~Negi, ``{Quaternion Octonion Reformulation of
  Grand Unified Theories},'' {\em International Journal of Theoretical
  Physics}, vol.~51, pp.~3228--3235, Oct. 2012.

\bibitem{Potter_2015}
F.~Potter, ``{CKM} and {PMNS} mixing matrices from discrete subgroups of
  {SU}(2),'' {\em Journal of Physics: Conference Series}, vol.~631, p.~012024,
  jul 2015.

\bibitem{Octon}
T.~Dray and C.~A. Manogue, {\em The geometry of the octonions}.
\newblock World Scientific, 2015.

\bibitem{Chanyal_2012}
B.~C. Chanyal, P.~S. Bisht, T.~Li, and O.~P.~S. Negi, ``Octonion quantum
  chromodynamics,'' {\em International Journal of Theoretical Physics},
  vol.~51, no.~11, p.~3410–3422, 2012.

\bibitem{Book_Wilson}
R.~Wilson, {\em The finite simple groups}.
\newblock Springer, 2009.

\bibitem{Holland_2003}
K.~Holland, P.~Minkowski, M.~Pepe, and U.-J. Wiese, ``{Exceptional confinement
  in G(2) gauge theory},'' {\em Nuclear Physics B}, vol.~668, p.~207–236, Sep
  2003.

\bibitem{2018IJMPA..3330005F}
C.~Furey, ``{A demonstration that electroweak theory can violate parity
  automatically (leptonic case)},'' {\em International Journal of Modern
  Physics A}, vol.~33, p.~1830005, Feb. 2018.

\bibitem{FUREY2015195}
C.~Furey, ``Charge quantization from a number operator,'' {\em Physics Letters
  B}, vol.~742, pp.~195 -- 199, 2015.

\bibitem{refId0F}
C.~Furey, ``{$SU(3)\times SU(2)\times U(1)(\times U(1))$ as a symmetry of
  division algebraic ladder operators},'' {\em Eur. Phys. J. C}, vol.~78,
  no.~5, p.~375, 2018.

\bibitem{PhysRevD.86.025024}
C.~Furey, ``Unified theory of ideals,'' {\em Phys. Rev. D}, vol.~86, p.~025024,
  Jul 2012.

\bibitem{Anastasiou_2014}
A.~Anastasiou, L.~Borsten, M.~J. Duff, L.~J. Hughes, and S.~Nagy, ``{Super
  Yang-Mills, division algebras and triality},'' {\em Journal of High Energy
  Physics}, vol.~2014, Aug 2014.

\bibitem{QFT4Math_Deligne}
P.~Deligne, {\em Quantum fields and strings: a course for mathematicians}.
\newblock American Mathematical Society; [Princeton, NJ]: Institute for
  Advanced Study, 1999.

\bibitem{Book_Superstring}
M.~M.~B. Green, J.~H. Schwarz, and E.~Witten, {\em Superstring theory. Volume
  1, Introduction}.
\newblock Cambridge University Press, 25th anniversary ed~ed., 2012.

\bibitem{Kugo:1982bn}
T.~Kugo and P.~K. Townsend, ``{Supersymmetry and the Division Algebras},'' {\em
  Nucl. Phys.}, vol.~B221, pp.~357--380, 1983.

\bibitem{Preitschopf:1995qf}
C.~R. Preitschopf, ``{Octonions and supersymmetry},'' in {\em {Gauge theories,
  applied supersymmetry, quantum gravity. Proceedings, Workshop, Leuven,
  Belgium, July 10-14, 1995}}, pp.~225--231, 1995.

\bibitem{g2manifold}
D.~D. Joyce, {\em Compact manifolds with special holonomy}.
\newblock Oxford University Press, 2000.

\bibitem{Becker_2015}
K.~Becker, M.~Becker, and D.~Robbins, ``M-theory and g2 manifolds,'' {\em
  Physica Scripta}, vol.~90, p.~118004, oct 2015.

\bibitem{culbert_2007}
C.~Culbert, ``{Cayley-Dickson algebras and loops},'' {\em Journal of
  Generalized Lie Theory and Applications}, vol.~01, 01 2007.

\bibitem{SMITH1995128}
J.~Smith, ``{A Left Loop on the 15-Sphere},'' {\em Journal of Algebra},
  vol.~176, no.~1, pp.~128 -- 138, 1995.

\bibitem{moreno1997zero}
R.~G. Moreno, ``{The zero divisors of the Cayley-Dickson algebras over the real
  numbers},'' {\em q-alg/9710013}, 1997.

\bibitem{barton2000magic}
C.~H. Barton and A.~Sudbery, ``Magic squares of lie algebras,'' {\em
  math/0001083}, 2000.

\bibitem{BARTON2003596}
C.~Barton and A.~Sudbery, ``Magic squares and matrix models of lie algebras,''
  {\em Advances in Mathematics}, vol.~180, no.~2, pp.~596 -- 647, 2003.

\bibitem{Cacciatori:2012cb}
S.~L. Cacciatori, B.~L. Cerchiai, and A.~Marrani, ``{Squaring the Magic},''
  {\em Adv. Theor. Math. Phys.}, vol.~19, pp.~923--954, 2015.

\bibitem{gillard2019cell8}
A.~B. Gillard and N.~G. Gresnigt, ``{The $C\ell(8)$ algebra of three fermion
  generations with spin and full internal symmetries},'' {\em
  arXiv:1906.05102}, 2019.

\bibitem{Brown67}
R.~B. Brown, ``{On Generalized Cayley-Dickson Algebras},'' {\em Pacific Journal
  of Mathematics}, vol.~20, no.~3, 1967.

\bibitem{EakinAutomorph}
P.~Eakin and A.~Sathaye, ``{On automorphisms and derivations of Cayley-Dickson
  algebras},'' {\em Journal of Algebra}, vol.~129, no.~2, pp.~263 -- 278, 1990.

\bibitem{triginta}
I.~Hentzel, ``{Identities For Algebras Obtained From The Cayley-Dickson
  Process},'' {\em Communications in Algebra}, vol.~29, 09 2000.

\bibitem{Kubo_2004}
J.~Kubo, H.~Okada, and F.~Sakamaki, ``{Higgs potential in a minimal $S_3$
  invariant extension of the standard model},'' {\em Physical Review D},
  vol.~70, Aug 2004.

\bibitem{Kubo_2005}
J.~Kubo, A.~Mondrag{\'{o}}n, M.~Mondrag{\'{o}}n,
  E.~Rodr{\'{\i}}guez-J{\'{a}}uregui, O.~F{\'{e}}lix-Beltr{\'{a}}n, and
  E.~Peinado, ``{A minimal $S_3$-invariant extension of the Standard Model},''
  {\em Journal of Physics: Conference Series}, vol.~18, pp.~380--384, jan 2005.

\bibitem{Mondrag_n_2007}
A.~Mondragón, M.~Mondragón, and E.~Peinado, ``{Lepton masses, mixings, and
  flavor-changing neutral currents in a minimal $S_3$-invariant extension of
  the standard model},'' {\em Physical Review D}, vol.~76, Oct 2007.

\bibitem{Gonz_lez_Canales_2012}
F.~González~Canales, A.~Mondragón, and M.~Mondragón, ``{The $S_3$ flavour
  symmetry: Neutrino masses and mixings},'' {\em Fortschritte der Physik},
  vol.~61, p.~546–570, Oct 2012.

\bibitem{Book_Aschbacher}
M.~Aschbacher, {\em Finite group theory}.
\newblock Cambridge University Press, 2nd~ed., 2000.

\bibitem{AdvModAlg}
J.~J. Rotman, {\em Advanced modern algebra}.
\newblock Providence, Rhode Island: American Mathematical Society, third~ed.,
  2015.

\bibitem{FondQCD}
T.~Muta, {\em Foundations of quantum chromodynamics: an introduction to
  perturbative methods in gauge theories}.
\newblock World Scientific, 3rd~ed., 2010.

\bibitem{Pepe_2006}
M.~Pepe, ``Confinement and the center of the gauge group,'' {\em Nuclear
  Physics B - Proceedings Supplements}, vol.~153, p.~207–214, Mar 2006.

\bibitem{Wipf:2013vp}
A.~Wipf, {\em {Statistical approach to quantum field theory}: {An
  introduction}}, vol.~864.
\newblock Springer, 2013.

\bibitem{Marzocca_2014}
D.~Marzocca and A.~Urbano, ``{Composite dark matter and LHC interplay},'' {\em
  Journal of High Energy Physics}, vol.~2014, Jul 2014.

\bibitem{Da_Rold_2019}
L.~Da~Rold and A.~N. Rossia, ``{The minimal simple composite Higgs model},''
  {\em Journal of High Energy Physics}, vol.~2019, Dec 2019.

\bibitem{Cacciapaglia_2019}
G.~Cacciapaglia, H.~Cai, A.~Deandrea, and A.~Kushwaha, ``{Composite Higgs and
  Dark Matter model in SU(6)/SO(6)},'' {\em Journal of High Energy Physics},
  vol.~2019, Oct 2019.

\bibitem{Coset}
A.~J. Macfarlane, ``{THE SPHERE S6 VIEWED AS A G2/SU(3) COSET SPACE},'' {\em
  International Journal of Modern Physics A}, vol.~17, no.~19, pp.~2595--2613,
  2002.

\bibitem{RevModPhys.34.1}
R.~E. Behrends, J.~Dreitlein, C.~Fronsdal, and W.~Lee, ``Simple groups and
  strong interaction symmetries,'' {\em Rev. Mod. Phys.}, vol.~34, pp.~1--40,
  Jan 1962.

\bibitem{Carone_2008}
C.~D. Carone and A.~Rastogi, ``Exceptional electroweak model,'' {\em Physical
  Review D}, vol.~77, Feb 2008.

\bibitem{PhysRevD.99.116024}
Z.~Dehghan and S.~Deldar, ``{Cho decomposition, Abelian gauge fixing, and
  monopoles in G(2) Yang-Mills theory},'' {\em Phys. Rev. D}, vol.~99,
  p.~116024, Jun 2019.

\bibitem{Pepe_2007}
M.~Pepe and U.-J. Wiese, ``Exceptional deconfinement in gauge theory,'' {\em
  Nuclear Physics B}, vol.~768, p.~21–37, Apr 2007.

\bibitem{Greensite_2007}
J.~Greensite, K.~Langfeld, .~Olejník, H.~Reinhardt, and T.~Tok, ``{Color
  screening, Casimir scaling, and domain structure in G(2) and SU(N) gauge
  theories},'' {\em Physical Review D}, vol.~75, Feb 2007.

\bibitem{Nejad_2014}
S.~H. Nejad and S.~Deldar, ``{Role of the SU(2) and SU(3) subgroups in
  observing confinement in the G(2) gauge group},'' {\em Physical Review D},
  vol.~89, Jan 2014.

\bibitem{Wellegehausen:2011jz}
B.~H. Wellegehausen, ``{Phase diagram of the G(2) Higgs model and G(2)-QCD},''
  {\em PoS}, vol.~LATTICE2011, p.~266, 2011.

\bibitem{Adhikary_2019}
A.~Adhikary, S.~Banerjee, R.~K. Barman, and B.~Bhattacherjee, ``{Resonant heavy
  Higgs searches at the HL-LHC},'' {\em Journal of High Energy Physics},
  vol.~2019, Sep 2019.

\bibitem{Arhrib_2014}
A.~Arhrib, P.~M. Ferreira, and R.~Santos, ``{Are there hidden scalars in LHC
  Higgs results?},'' {\em Journal of High Energy Physics}, vol.~2014, Mar 2014.

\bibitem{Banerjee_2018}
A.~Banerjee, G.~Bhattacharyya, N.~Kumar, and T.~S. Ray, ``{Constraining
  composite Higgs models using LHC data},'' {\em Journal of High Energy
  Physics}, vol.~2018, Mar 2018.

\bibitem{Gonz_lez_Canales_2013}
F.~González~Canales, A.~Mondragón, M.~Mondragón, U.~J. Saldaña~Salazar, and
  L.~Velasco-Sevilla, ``{Quark sector of $S_3$ models: Classification and
  comparison with experimental data},'' {\em Physical Review D}, vol.~88, Nov
  2013.

\bibitem{Lucini_2004}
B.~Lucini, M.~Teper, and U.~Wenger, ``{The high temperature phase transition in
  SU(N) gauge theories},'' {\em Journal of High Energy Physics}, vol.~2004,
  p.~061–061, Jan 2004.

\bibitem{Nada:2015aia}
A.~Nada, ``{Universal aspects in the equation of state for Yang-Mills
  theories},'' {\em PoS}, vol.~EPS-HEP2015, p.~373, 2015.

\bibitem{Lucini_2014}
B.~Lucini and M.~Panero, ``{Introductory lectures to large- QCD phenomenology
  and lattice results},'' {\em Progress in Particle and Nuclear Physics},
  vol.~75, p.~1–40, Mar 2014.

\bibitem{Teper:2009uf}
M.~Teper, ``{Large N and confining flux tubes as strings - a view from the
  lattice},'' {\em Acta Phys. Polon.}, vol.~B40, pp.~3249--3320, 2009.

\bibitem{Panero_2009}
M.~Panero, ``{Thermodynamics of the QCD Plasma and the Large-N Limit},'' {\em
  Physical Review Letters}, vol.~103, Dec 2009.

\bibitem{Cossu_2007}
G.~Cossu, M.~D'Elia, A.~D. Giacomo, B.~Lucini, and C.~Pica, ``G2 gauge theory
  at finite temperature,'' {\em Journal of High Energy Physics}, vol.~2007,
  p.~100–100, Oct 2007.

\bibitem{vonSmekal:2013qqa}
L.~von Smekal, B.~H. Wellegehausen, A.~Maas, and A.~Wipf, ``{$G_2$-QCD:
  Spectroscopy and the phase diagram at zero temperature and finite density},''
  {\em PoS}, vol.~LATTICE2013, p.~186, 2014.

\bibitem{thermoG22014}
M.~Bruno, M.~Caselle, M.~Panero, and R.~Pellegrini, ``{Exceptional
  thermodynamics: The equation of state of G(2) gauge theory},'' {\em Journal
  of High Energy Physics}, vol.~2015, 09 2014.

\bibitem{Cutting_2018}
D.~Cutting, M.~Hindmarsh, and D.~J. Weir, ``{Gravitational waves from vacuum
  first-order phase transitions: From the envelope to the lattice},'' {\em
  Physical Review D}, vol.~97, Jun 2018.

\bibitem{PhysRevLett.115.181101}
P.~Schwaller, ``Gravitational waves from a dark phase transition,'' {\em Phys.
  Rev. Lett.}, vol.~115, p.~181101, Oct 2015.

\bibitem{Zhou_2020}
R.~Zhou, J.~Yang, and L.~Bian, ``{Gravitational waves from first-order phase
  transition and domain wall},'' {\em Journal of High Energy Physics},
  vol.~2020, Apr 2020.

\bibitem{2021JHEP...05..160Z}
Z.~{Zhang}, C.~{Cai}, X.-M. {Jiang}, Y.-L. {Tang}, Z.-H. {Yu}, and H.-H.
  {Zhang}, ``{Phase transition gravitational waves from pseudo-Nambu-Goldstone
  dark matter and two Higgs doublets},'' {\em Journal of High Energy Physics},
  vol.~2021, p.~160, May 2021.

\bibitem{masses2014}
B.~Wellegehausen, A.~Maas, A.~Wipf, and L.~Von~Smekal, ``{Hadron masses and
  baryonic scales in $G_2$-QCD at finite density},'' {\em Physical Review D},
  vol.~89, p.~056007, 03 2014.

\bibitem{Hajizadeh_2017}
O.~Hajizadeh and A.~Maas, ``{Constructing a neutron star from the lattice in
  G2-QCD},'' {\em The European Physical Journal A}, vol.~53, Oct 2017.

\bibitem{Juknevich_2009}
J.~Juknevich, D.~Melnikov, and M.~Strassler, ``{A pure-glue hidden valley I.
  States and decays},'' {\em Journal of High Energy Physics}, vol.~2009,
  p.~055–055, Jul 2009.

\bibitem{Juknevich:2009gg}
J.~E. Juknevich, ``{Pure-glue hidden valleys through the Higgs portal},'' {\em
  JHEP}, vol.~08, p.~121, 2010.

\bibitem{Boddy_2014}
K.~K. Boddy, J.~L. Feng, M.~Kaplinghat, and T.~M. Tait, ``Self-interacting dark
  matter from a non-abelian hidden sector,'' {\em Physical Review D}, vol.~89,
  Jun 2014.

\bibitem{Yamanaka:2014pva}
N.~Yamanaka, S.~Fujibayashi, S.~Gongyo, and H.~Iida, ``{Dark matter in the
  hidden gauge theory},'' {\em arXiv:1411.2172}, 2014.

\bibitem{Klinkhamer_2010}
F.~R. Klinkhamer, ``Gluon condensate, modified gravity, and the accelerating
  universe,'' {\em Physical Review D}, vol.~81, Feb 2010.

\bibitem{Ballesteros_2017}
G.~Ballesteros, A.~Carmona, and M.~Chala, ``Exceptional composite dark
  matter,'' {\em The European Physical Journal C}, vol.~77, Jul 2017.

\bibitem{Koorambas:2013una}
E.~Koorambas, ``{Vector Gauge Boson Dark Matter for the $SU(N)$ Gauge Group
  Model},'' {\em Int. J. Theor. Phys.}, vol.~52, pp.~4374--4388, 2013.

\bibitem{Yukawa_2012}
E.~Yukawa and M.~Ueda, ``{Hydrodynamic description of spin-1 Bose-Einstein
  condensates},'' {\em Physical Review A}, vol.~86, Dec 2012.

\bibitem{PhysRevD.97.075029}
L.~Forestell, D.~E. Morrissey, and K.~Sigurdson, ``Cosmological bounds on
  non-abelian dark forces,'' {\em Phys. Rev. D}, vol.~97, p.~075029, Apr 2018.

\bibitem{SONI2017379}
A.~Soni and Y.~Zhang, ``{Gravitational waves from SU(N) glueball dark
  matter},'' {\em Physics Letters B}, vol.~771, pp.~379 -- 384, 2017.

\bibitem{PhysRevLett.77.2622}
G.~B. West, ``Theorem on the lightest glueball state,'' {\em Phys. Rev. Lett.},
  vol.~77, pp.~2622--2625, Sep 1996.

\bibitem{2015PhLB..746..430A}
A.~{Ahriche}, K.~L. {McDonald}, S.~{Nasri}, and T.~{Toma}, ``{A model of
  neutrino mass and dark matter with an accidental symmetry},'' {\em Physics
  Letters B}, vol.~746, pp.~430--435, June 2015.

\bibitem{Bernal_2016}
N.~Bernal and X.~Chu, ``{$Z_2$ SIMP dark matter},'' {\em Journal of Cosmology
  and Astroparticle Physics}, vol.~2016, p.~006–006, Jan 2016.

\bibitem{Branco_2012}
G.~Branco, P.~Ferreira, L.~Lavoura, M.~Rebelo, M.~Sher, and J.~P. Silva,
  ``Theory and phenomenology of two-higgs-doublet models,'' {\em Physics
  Reports}, vol.~516, p.~1–102, Jul 2012.

\bibitem{Yaguna_2020}
C.~E. Yaguna and s.~Zapata, ``{Multi-component scalar dark matter from a $Z_N$
  symmetry: a systematic analysis},'' {\em Journal of High Energy Physics},
  vol.~2020, Mar 2020.

\bibitem{Bai_2010}
Y.~Bai and R.~J. Hill, ``{Weakly interacting stable hidden sector pions},''
  {\em Physical Review D}, vol.~82, Dec 2010.

\bibitem{Rinaldi_2018}
M.~Rinaldi and V.~Vento, ``Scalar and tensor glueballs as gravitons,'' {\em The
  European Physical Journal A}, vol.~54, Sep 2018.

\bibitem{scalarvectorgravity}
I.~Quiros, ``Selected topics in scalar-tensor theories and beyond,'' {\em
  International Journal of Modern Physics D}, vol.~28, no.~07, p.~1930012,
  2019.

\bibitem{Akrami_2015}
Y.~Akrami, S.~Hassan, F.~Könnig, A.~Schmidt-May, and A.~R. Solomon, ``Bimetric
  gravity is cosmologically viable,'' {\em Physics Letters B}, vol.~748,
  p.~37–44, Sep 2015.

\bibitem{PhysRevD.94.084055}
E.~Babichev, L.~Marzola, M.~Raidal, A.~Schmidt-May, F.~Urban, H.~Veerm\"ae, and
  M.~von Strauss, ``Bigravitational origin of dark matter,'' {\em Phys. Rev.
  D}, vol.~94, p.~084055, Oct 2016.

\bibitem{Babichev_2016}
E.~Babichev, L.~Marzola, M.~Raidal, A.~Schmidt-May, F.~Urban, H.~Veermäe, and
  M.~v. Strauss, ``Heavy spin-2 dark matter,'' {\em Journal of Cosmology and
  Astroparticle Physics}, vol.~2016, p.~016–016, Sep 2016.

\bibitem{Giacosa:2017eqy}
F.~Giacosa, ``{Heavy Glueballs: Status and Large-$N_{\rm c}$ Widths
  Estimate},'' {\em Acta Phys. Polon. Supp.}, vol.~10, pp.~1021--1027, 2017.

\bibitem{Lucini:2014paa}
B.~Lucini, ``{Glueballs from the Lattice},'' {\em PoS}, vol.~QCD-TNT-III,
  p.~023, 2013.

\bibitem{daRocha:2017cxu}
R.~da~Rocha, ``{Dark SU(N) glueball stars on fluid branes},'' {\em Phys. Rev.},
  vol.~D95, no.~12, p.~124017, 2017.

\bibitem{Aghanim:2018eyx}
N.~Aghanim {\em et~al.}, ``{Planck 2018 results. VI. Cosmological
  parameters},'' {\em arXiv:1807.06209}, 2018.

\bibitem{Bernal_2017}
N.~Bernal, M.~Heikinheimo, T.~Tenkanen, K.~Tuominen, and V.~Vaskonen, ``{The
  dawn of FIMP Dark Matter: A review of models and constraints},'' {\em
  International Journal of Modern Physics A}, vol.~32, p.~1730023, Sep 2017.

\bibitem{Hall_2010}
L.~J. Hall, K.~Jedamzik, J.~March-Russell, and S.~M. West, ``Freeze-in
  production of fimp dark matter,'' {\em Journal of High Energy Physics},
  vol.~2010, Mar 2010.

\bibitem{Bernal_2019}
N.~Bernal, C.~Cosme, and T.~Tenkanen, ``{Phenomenology of self-interacting dark
  matter in a matter-dominated universe},'' {\em The European Physical Journal
  C}, vol.~79, Jan 2019.

\bibitem{Yamanaka:2019yek}
N.~Yamanaka, H.~Iida, A.~Nakamura, and M.~Wakayama, ``{Glueball scattering
  cross section in lattice SU(2) Yang-Mills theory},'' {\em Phys. Rev. D},
  vol.~102, no.~5, p.~054507, 2020.

\bibitem{Soni:2016gzf}
A.~Soni and Y.~Zhang, ``{Hidden SU(N) Glueball Dark Matter},'' {\em Phys. Rev.
  D}, vol.~93, no.~11, p.~115025, 2016.

\bibitem{Enqvist_2018}
K.~Enqvist, R.~J. Hardwick, T.~Tenkanen, V.~Vennin, and D.~Wands, ``A novel way
  to determine the scale of inflation,'' {\em Journal of Cosmology and
  Astroparticle Physics}, vol.~2018, pp.~006--006, feb 2018.

\bibitem{Bernal_2016bis}
N.~Bernal, X.~Chu, C.~Garcia-Cely, T.~Hambye, and B.~Zaldivar, ``Production
  regimes for self-interacting dark matter,'' {\em Journal of Cosmology and
  Astroparticle Physics}, vol.~2016, p.~018–018, Mar 2016.

\bibitem{Heikinheimo_2017}
M.~Heikinheimo, T.~Tenkanen, K.~Tuominen, and V.~Vaskonen, ``{Observational
  Constraints on Decoupled Hidden Sectors},'' {\em Phys. Rev. D}, vol.~94,
  no.~6, p.~063506, 2016.
\newblock [Erratum: Phys.Rev.D 96, 109902 (2017)].

\bibitem{Choi_2017}
S.-M. Choi, H.~M. Lee, and M.-S. Seo, ``{Cosmic abundances of SIMP dark
  matter},'' {\em Journal of High Energy Physics}, vol.~2017, Apr 2017.

\bibitem{Forestell:2016qhc}
L.~Forestell, D.~E. Morrissey, and K.~Sigurdson, ``{Non-Abelian Dark Forces and
  the Relic Densities of Dark Glueballs},'' {\em Phys. Rev.}, vol.~D95, no.~1,
  p.~015032, 2017.

\bibitem{Bhattacharya_2020}
S.~Bhattacharya, P.~Ghosh, and S.~Verma, ``Simpler realisation of scalar dark
  matter,'' {\em Journal of Cosmology and Astroparticle Physics}, vol.~2020,
  p.~040–040, Jan 2020.

\bibitem{Choi:2020ara}
S.-M. Choi, J.~Kim, H.~M. Lee, and B.~Zhu, ``{Connecting between inflation and
  dark matter in models with gauged Z$_{3}$ symmetry},'' {\em JHEP}, vol.~06,
  p.~135, 2020.

\bibitem{2017arXiv170401804S}
B.~{Samir Acharya}, M.~{Fairbairn}, and E.~{Hardy}, ``{Glueball dark matter in
  non-standard cosmologies},'' {\em arXiv e-prints}, p.~arXiv:1704.01804, Apr.
  2017.

\bibitem{Allahverdi:2002nb}
R.~Allahverdi and M.~Drees, ``{Production of massive stable particles in
  inflaton decay},'' {\em Phys. Rev. Lett.}, vol.~89, p.~091302, 2002.

\bibitem{Almeida:2018oid}
J.~P.~B. Almeida, N.~Bernal, J.~Rubio, and T.~Tenkanen, ``{Hidden inflation
  dark matter},'' {\em JCAP}, vol.~03, p.~012, 2019.

\bibitem{delaMacorra:2012sb}
A.~de~la Macorra, ``{Dark Matter from the Inflaton Field},'' {\em Astropart.
  Phys.}, vol.~35, pp.~478--484, 2012.

\bibitem{Heurtier:2019eou}
L.~Heurtier and F.~Huang, ``{Inflaton portal to a highly decoupled EeV dark
  matter particle},'' {\em Phys. Rev. D}, vol.~100, no.~4, p.~043507, 2019.

\bibitem{Mirza_2011}
B.~Mirza and H.~Mohammadzadeh, ``Condensation of an ideal gas obeying
  non-abelian statistics,'' {\em Physical Review E}, vol.~84, Sep 2011.

\bibitem{Harko_2011}
T.~Harko, ``{Cosmological dynamics of dark matter Bose-Einstein
  condensation},'' {\em Physical Review D}, vol.~83, Jun 2011.

\bibitem{Sharma:2018ydn}
A.~Sharma, J.~Khoury, and T.~Lubensky, ``{The Equation of State of Dark Matter
  Superfluids},'' {\em JCAP}, vol.~1905, no.~05, p.~054, 2019.

\bibitem{Ferreira_2019}
E.~G. Ferreira, G.~Franzmann, J.~Khoury, and R.~Brandenberger, ``Unified
  superfluid dark sector,'' {\em Journal of Cosmology and Astroparticle
  Physics}, vol.~2019, p.~027–027, Aug 2019.

\bibitem{Axion}
M.~M. Kuster, B.~B. Beltran, and G.~G. Raffelt, {\em Axions: theory, cosmology,
  and experimental searches}.
\newblock Berlin: Springer, 2010.

\bibitem{Book_BEC}
S.~D.~W. Proukakis, Nick~P. and P.~B. Littlewood, {\em {Universal themes of
  Bose-Einstein condensation}}.
\newblock Cambridge University Press, 2017.

\bibitem{B_hmer_2007}
C.~G. Balhmer and T.~Harko, ``{Can dark matter be a Bose-Einstein
  condensate?},'' {\em Journal of Cosmology and Astroparticle Physics},
  vol.~2007, pp.~025--025, jun 2007.

\bibitem{Chavanis:2016ial}
P.-H. Chavanis, ``{Dissipative self-gravitating Bose-Einstein condensates with
  arbitrary nonlinearity as a model of dark matter halos},'' {\em Eur. Phys. J.
  Plus}, vol.~132, no.~6, p.~248, 2017.

\bibitem{Sanchis-Gual:2019ljs}
N.~Sanchis-Gual, F.~Di~Giovanni, M.~Zilhão, C.~Herdeiro, P.~Cerdá-Durán, J.~A.
  Font, and E.~Radu, ``{Nonlinear Dynamics of Spinning Bosonic Stars: Formation
  and Stability},'' {\em Phys. Rev. Lett.}, vol.~123, no.~22, p.~221101, 2019.

\bibitem{Eby:2015hsq}
J.~Eby, C.~Kouvaris, N.~G. Nielsen, and L.~C.~R. Wijewardhana, ``{Boson Stars
  from Self-Interacting Dark Matter},'' {\em JHEP}, vol.~02, p.~028, 2016.

\bibitem{Liebling_2017}
S.~L. Liebling and C.~Palenzuela, ``Dynamical boson stars,'' {\em Living
  Reviews in Relativity}, vol.~20, Nov 2017.

\bibitem{Brito:2015pxa}
R.~Brito, V.~Cardoso, C.~A.~R. Herdeiro, and E.~Radu, ``{Proca stars:
  Gravitating Bose–Einstein condensates of massive spin 1 particles},'' {\em
  Phys. Lett.}, vol.~B752, pp.~291--295, 2016.

\bibitem{Landea_2016}
I.~S. Landea and F.~García, ``{Charged Proca stars},'' {\em Physical Review D},
  vol.~94, Nov 2016.

\bibitem{Minamitsuji_2018}
M.~Minamitsuji, ``Vector boson star solutions with a quartic order
  self-interaction,'' {\em Physical Review D}, vol.~97, May 2018.

\bibitem{Chavanis_2012}
P.-H. Chavanis and T.~Harko, ``{Bose-Einstein condensate general relativistic
  stars},'' {\em Physical Review D}, vol.~86, Sep 2012.

\bibitem{Cardoso_2019}
V.~Cardoso and P.~Pani, ``{Testing the nature of dark compact objects: a status
  report},'' {\em Living Reviews in Relativity}, vol.~22, Jul 2019.

\bibitem{Hertzberg_2021}
M.~P. Hertzberg, F.~Rompineve, and J.~Yang, ``{Decay of boson stars with
  application to glueballs and other real scalars},'' {\em Physical Review D},
  vol.~103, Jan 2021.

\bibitem{Calcino:2018mwh}
J.~Calcino, J.~Garcia-Bellido, and T.~M. Davis, ``{Updating the MACHO fraction
  of the Milky Way dark halowith improved mass models},'' {\em Mon. Not. Roy.
  Astron. Soc.}, vol.~479, no.~3, pp.~2889--2905, 2018.

\bibitem{Brandt:2016aco}
T.~D. Brandt, ``{Constraints on MACHO Dark Matter from Compact Stellar Systems
  in Ultra-Faint Dwarf Galaxies},'' {\em Astrophys. J. Lett.}, vol.~824, no.~2,
  p.~L31, 2016.

\bibitem{Harvey_2019}
D.~Harvey, A.~Robertson, R.~Massey, and I.~G. McCarthy, ``Observable tests of
  self-interacting dark matter in galaxy clusters: Bcg wobbles in a constant
  density core,'' {\em Monthly Notices of the Royal Astronomical Society},
  vol.~488, p.~1572–1579, Jul 2019.

\bibitem{Bernal_2020}
N.~Bernal, X.~Chu, S.~Kulkarni, and J.~Pradler, ``Self-interacting dark matter
  without prejudice,'' {\em Physical Review D}, vol.~101, Mar 2020.

\bibitem{Robertson_2016}
A.~Robertson, R.~Massey, and V.~Eke, ``What does the bullet cluster tell us
  about self-interacting dark matter?,'' {\em Monthly Notices of the Royal
  Astronomical Society}, vol.~465, p.~569–587, Oct 2016.

\bibitem{Ohnishi:2011jv}
A.~Ohnishi, H.~Ueda, T.~Z. Nakano, M.~Ruggieri, and K.~Sumiyoshi,
  ``{Possibility of QCD critical point sweep during black hole formation},''
  {\em Phys. Lett.}, vol.~B704, pp.~284--290, 2011.

\bibitem{Critelli:2017oub}
R.~Critelli, J.~Noronha, J.~Noronha-Hostler, I.~Portillo, C.~Ratti, and
  R.~Rougemont, ``{Critical point in the phase diagram of primordial
  quark-gluon matter from black hole physics},'' {\em Phys. Rev.}, vol.~D96,
  no.~9, p.~096026, 2017.

\end{thebibliography}
\bibliographystyle{ieeetr}

%
%

\end{document}